\documentclass[aps,pre,showpacs,showkeys,onecolumn,superscriptaddress,amsmath,amssymb]{revtex4}
\usepackage[dvips]{graphicx}

\newcommand{\ie}{\emph{i.e.}, }
\newcommand{\TT}{\mathrm{T}}
\newcommand{\Tr}{\mathrm{Tr}}
\newcommand{\Det}{\mathrm{Det}}
\newcommand{\dd}{\mathrm{d}}
\newcommand{\DD}{\mathrm{D}}

\newcommand{\im}{\mathrm{Im}}
\newcommand{\ii}{\mathrm{i}}
\newcommand{\ee}{\mathrm{e}}
\newcommand{\diag}{\mathrm{diag}}
\newcommand{\la}{\left\langle}
\newcommand{\ra}{\right\rangle}

\begin{document}

\title{Applying Free Random Variables\\to Random Matrix Analysis of Financial Data\\Part I: A Gaussian Case}
\author{Zdzis\l{}aw Burda}
\email{zdzislaw.burda@uj.edu.pl}
\affiliation{Marian Smoluchowski Institute of Physics and Mark Kac Complex Systems Research Centre, Jagiellonian University, Reymonta 4, 30--059 Krak\'{o}w, Poland}
\author{Andrzej Jarosz}
\email{andrzej.jarosz@clico.pl}
\affiliation{Clico Ltd., Oleandry 2, 30--063 Krak\'{o}w, Poland}
\author{Jerzy Jurkiewicz}
\email{jerzy.jurkiewicz@uj.edu.pl}
\affiliation{Marian Smoluchowski Institute of Physics and Mark Kac Complex Systems Research Centre, Jagiellonian University, Reymonta 4, 30--059 Krak\'{o}w, Poland}
\author{Maciej A. Nowak}
\email{nowak@th.if.uj.edu.pl}
\affiliation{Marian Smoluchowski Institute of Physics and Mark Kac Complex Systems Research Centre, Jagiellonian University, Reymonta 4, 30--059 Krak\'{o}w, Poland}
\author{G\'{a}bor Papp}
\email{pg@ludens.elte.hu}
\affiliation{Institute for Physics, E\"{o}tv\"{o}s University, 1518 Budapest, Hungary}
\author{Ismail Zahed}
\email{zahed@zahed.physics.sunysb.edu}
\affiliation{Department of Physics and Astronomy, SUNY Stony Brook, NY 11794, USA}

\date{\today}

\setlength{\parindent}{4ex}
\setlength{\parskip}{1.5ex plus 0.5ex minus 0.2ex}
\topmargin=0cm

\begin{abstract}
We apply the concept of free random variables to doubly correlated (Gaussian) Wishart random matrix models, appearing for example in a multivariate analysis of financial time series, and displaying both inter--asset cross--covariances and temporal auto--covariances. We give a comprehensive introduction to the rich financial reality behind such models. We explain in an elementary way the main techniques of the free random variables calculus, with a view to promote them in the quantitative finance community. We apply our findings to tackle several financially relevant problems, such as of an universe of assets displaying exponentially decaying temporal covariances, or the exponentially weighted moving average, both with an arbitrary structure of cross--covariances.
\end{abstract}


\maketitle


\section{Introduction}
\label{s:Introduction}


\subsection{Financial Cross--Correlations and Auto--Correlations From Gaussian Random Matrix Theory}
\label{ss:FinancialCrossCorrelationsAndAutoCorrelationsFromGaussianRandomMatrixTheory}


\subsubsection{Financial Correlations and Portfolio Optimization}
\label{sss:FinancialCorrelationsAndPortfolioOptimization}

\emph{Cross--correlations} between assets traded on the markets play a critical role in the practice of modern--day financial institutions. For example, correlated moves of assets diminish a possibility of optimal diversification of investment portfolios, and precise knowledge of these correlations is fundamental for optimal capital allocation: in the classical Markowitz mean--variance optimization theory~\cite{Markowitz1952}, all the correlations must be perfectly known. Not only so, but a future \emph{forecast} of the correlations would be highly desirable. However, the information about cross--correlations and their temporal dynamics is typically inferred from historical data, stored in the memories of computers (usually in the form of large matrices); the material decoded from past time series is inevitably marred by measurement noise, and it is a constant challenge to unravel signal from noise.

One may argue that from the point of view of practical portfolio optimization, cross--correlations are only a ``second--order effect,'' since they represent fluctuations around a certain trend described by the average returns of assets. Now it is well--known that the future returns are to a great extent impossible to determine from the historical returns (one would need to take a very long time series to account for volatilities, which would manifest a highly non--stationary nature of the returns' distribution), and are thus subject to individual assessment by managers. We will disregard this problem in the present analysis, and assume that investors have some expectations of the future returns, recalling at the same time that volatilities and cross--correlations should be much more stable in time (due to the phenomenon of heteroscedasticity, \ie time--dependence of volatility), hence allowing some level of predictability of their future values from historical data, once the measurement noise has been properly cleaned.

Another reservation (see for example~\cite{GalluccioBouchaudPotters1998}) may be that optimization leads to a solution for the portfolio weights dependent on the covariance matrix in a stable way (in the sense that a small modification of the covariances leads to only a small change in the portfolio composition and risk) only in simplest cases of linear constraints on the weights, while for non--linear constrains (present for example on futures markets) the problem becomes equivalent to finding the energy minima of a spin glass system, and there is an exponentially large (in the number of assets) number of these local minima with an unstable (chaotic) dependence on the covariance matrix. Again, we will restrict our interest to portfolios with linear constraints on the weights, in which case the estimation noise of the covariance matrix is already troublesome enough for special techniques of its elimination to be unavoidable.

Keeping these limitations in mind, we conclude that correlations between assets necessarily have to be included in any risk analysis (portfolio optimization, option pricing). Even more importantly than investment purposes, one should investigate the covariance matrix out of a theoretical motivation, as its structure mirrors the interdependencies of companies, as well as possesses a non--trivial temporal dynamics; both qualities crucial for better understanding of the underlying mechanisms governing the behavior of financial markets.


\subsubsection{The Gaussian Approximation}
\label{sss:TheGaussianApproximation}

The primary way to describe not only the volatilities but also cross--correlations in a universe of some $N$ assets is through the two--point covariance function,
\begin{equation}\label{eq:CovarianceFunctionDefinition}
\mathcal{C}_{i a , j b} \equiv \la R_{i a} R_{j b} \ra .
\end{equation}
We define \smash{$r_{i a} \equiv \log S_{i , a + 1} - \log S_{i a} \approx ( S_{i , a + 1} - S_{i a} ) / S_{i a}$} to be the return of an asset $i = 1 , \ldots , N$ over a time interval $a = 1 , \ldots , T$, \ie (approximately) the relative change of the asset's price \smash{$S_{i a}$} between the moments of time $a \delta t$ and $( a + 1 ) \delta t$, where $\delta t$ is some elementary time step. (We disregard the known small ``leverage effect'' of anomalous skewness in the distribution of the returns: that the price increments, rather than the returns, behave as additive random variables.) Moreover, we denote \smash{$R_{i a} \equiv r_{i a} - \la r_{i a} \ra$}, which describe the fluctuations (with zero mean) of the returns around the trend, and collect them into a rectangular $N \times T$ matrix $\mathbf{R}$. The average $\langle \ldots \rangle$ is understood as taken according to some probability distribution whose functional shape is stable over time, but whose parameters may be time--dependent.

In this paper, we will employ a very simplified form of the two--point covariance function (\ref{eq:CovarianceFunctionDefinition}), namely with cross--covariances and auto--covariances factorized and non--random,
\begin{equation}\label{eq:CovarianceFunctionFactorization}
\mathcal{C}_{i a , j b} = C_{i j} A_{a b}
\end{equation}
(we will collect these coefficients into an $N \times N$ cross--covariance matrix $\mathbf{C}$ and a $T \times T$ auto--covariance matrix $\mathbf{A}$; both are taken symmetric and positive--definite). We will discover that the matrix of ``temporal covariances'' $\mathbf{A}$ is a way to model two temporal effects: the (weak, short--memory) lagged correlations between the returns (see par.~\ref{sss:LaggedCorrelationsBetweenTheReturns}), as well as the (stronger, long--memory) lagged correlations between the volatilities (heteroscedasticity; par.~\ref{sss:Heteroscedasticity}). On the other hand, the matrix of cross--covariances (``spatial covariances,'' using a more physical language) $\mathbf{C}$ models the hidden factors affecting the assets, thereby reflecting the structure of mutual dependencies of the market companies (par.~\ref{sss:FactorModels}). Importantly, both contributions are decoupled: the temporal dependence of the distribution of each asset is the same, and the structure of cross--correlations does not evolve in time; this is quite a crude approximation. Also, these are all fixed numbers; only in a subsequent work do we plan to explore another known way (a ``random parametric deformation'') of modeling temporal dependence of cross--covariances, that is, considering $\mathbf{C}$ to be a random matrix of a given probability distribution.

For our approach to be valid, both covariance matrices obviously must be finite. In fact, the current article deals exclusively with the multivariate Gaussian distribution for the assets' returns which displays the two--point covariances (\ref{eq:CovarianceFunctionFactorization}),
$$
P_{\mathrm{c.G.}} ( \mathbf{R} ) \DD \mathbf{R} = \frac{1}{\mathcal{N}_{\mathrm{c.G.}}} \exp \left( - \frac{1}{2} \sum_{i , j = 1}^{N} \sum_{a , b = 1}^{T} R_{i a} \left[ \mathbf{C}^{- 1} \right]_{i j} R_{j b} \left[ \mathbf{A}^{- 1} \right]_{b a} \right) \DD \mathbf{R} =
$$
\begin{equation}\label{eq:CorrelatedGaussian}
= \frac{1}{\mathcal{N}_{\mathrm{c.G.}}} \exp \left( - \frac{1}{2} \Tr \mathbf{R}^{\TT} \mathbf{C}^{- 1} \mathbf{R} \mathbf{A}^{- 1} \right) \DD \mathbf{R} ,
\end{equation}
where the normalization constant \smash{$\mathcal{N}_{\mathrm{c.G.}} = ( 2 \pi )^{N T / 2} ( \Det \mathbf{C} )^{T / 2} ( \Det \mathbf{A} )^{N / 2}$}, and the integration measure \smash{$\DD \mathbf{R} \equiv \prod_{i , a} \dd R_{i a}$}; the letters ``c.G.'' stand for ``correlated Gaussian,'' and the expectation map w.r.t. this distribution will be denoted by \smash{$\langle \ldots \rangle_{\mathrm{c.G.}}$}, while ``\smash{$^{\TT}$}'' denotes matrix transposition. In this case, the covariance matrices are finite, and moreover they are sufficient to fully characterize the dependencies of the \smash{$R_{i a}$}'s. However, in more realistic situations, such as of financially relevant distributions having heavy power--law tails (with an exponent $\mu$), the two--point covariance (in particular $\mathbf{C}$) may not exist. The precise answer boils down to how these correlated non--Gaussian distributions are defined (one can exploit a variety of methods: a linear model, a copula, a random deformation of some kind, a method based on freeness, on radial measures, etc.; see for example~\cite{BouchaudPotters2003}, sections 9.2, 12.2.3, 12.2.4, and~\cite{BurdaGorlichWaclaw2006,BiroliBouchaudPotters2007,BurdaJurkiewicz2009}), and we will postpone this discussion to forthcoming communications. Let us just mention that in some cases (such as a linear model of L\'{e}vy stable variables, \ie with $\mu < 2$), $\mathbf{C}$ diverges, and another measure of covariance should be devised (for example, the ``tail covariance,'' which quantifies the amplitude of the tail and asymmetry of the power--law product variable \smash{$R_{i a} R_{j b}$}); even for $\mu > 2$, when $\mathbf{C}$ exists, it is informative (especially from the point of view of portfolio optimization in the sense of minimizing value--at--risk) to use the tail covariance, as it focuses on large negative events. However, we henceforth restrict our attention to only the correlated Gaussian distribution (\ref{eq:CorrelatedGaussian}); on this simplest (and admittedly quite distant from reality) example we wish to advocate our approach, with a view to further generalize it to heavy--tailed distributions.


\subsubsection{Free Random Variables}
\label{sss:FreeRandomVariables}

A decade ago, Bouchaud \emph{et al.} and Stanley \emph{et al.}~\cite{LalouxCizeauBouchaudPotters1999,PlerouGopikrishnanRosenowAmaralStanley1999} suggested the use of Gaussian \emph{random matrix theory} (RMT) for addressing the issue of noise in financial correlation matrices. Since then, a number of results regarding the quantification of noise in financial covariances have been derived using this tool~\cite{LalouxCizeauPottersBouchaud2000,PlerouGopikrishnanRosenowAmaralGuhrStanley2002,DrozdzKwapienGrummerRufSpeth2001,LilloMantegna2003,RepetowiczRichmond2004,UtsugiInoOshikawa2003,PafkaKondor2002,PafkaKondor2003,PafkaKondor2004,PappPafkaNowakKondor2005,GuhrKalber2003,MalevergneSornette2004,BurdaGorlichJaroszJurkiewicz2004,BurdaJurkiewicz2004,BurdaJurkiewiczWaclaw2005-1,BurdaGorlichJurkiewiczWaclaw2006,BurdaJurkiewiczWaclaw2005-2}, some of which maybe of relevance to risk management.

In this work, we would like to advertise the concepts of the \emph{free random variables} (FRV) calculus as a powerful alternative to standard random matrix theory, both for Gaussian and non--Gaussian noise. FRV may be thought of as an abstract non--commutative generalization of the classical (commutative) probability calculus, \ie a mathematical framework for dealing with random variables which do not commute, examples of which are random matrices. (Indeed, FRV was initiated by Voiculescu \emph{et al.} and Speicher~\cite{VoiculescuDykemaNica1992,Speicher1994} as a rather abstract approach to von Neumann algebras, but it has a concrete realization in the context of RMT, since large random matrices can be regarded as free random variables.) Its centerpiece is a mathematical construction of the notion of \emph{freeness}, which is a non--commutative counterpart of classical independence of random variables. As such, it allows for extending many classical results founded upon the properties of independence into the non--commutative (random matrix) realm, particularly the algorithms of addition and multiplication of random variables, or the ideas of stability, infinite divisibility, \emph{etc.} This introduces a new quality into RMT, which simplifies, both conceptually and technically, many random matrix calculations, especially in the macroscopic limit (the bulk limit, \ie random matrices of infinite size), which is of main interest in practical problems.

Several years ago, we suggested that FRV is very useful for addressing a much larger class of noise (L\'{e}vy) in the context of financial covariance matrices in a way that is succinct and mostly algebraic~\cite{BurdaJanikJurkiewiczNowakPappZahed2002,BurdaJurkiewiczNowakPappZahed2003,BurdaJurkiewiczNowakPappZahed2004,BurdaJurkiewiczNowakPappZahed2001,BurdaJurkiewiczNowak2003,BurdaJurkiewiczNowakPappZahed2007}. These results have now seen further applications to financial covariances~\cite{NowakUnpublished2004,PottersBouchaudLaloux2005,BouchaudPotters2009} and macroeconomy~\cite{BouchaudLalouxAugustaMiceliPotters2007}. Also, FRV has been already applied to a number of problems ranging from physics~{\cite{NeuSpeicher1994,GopakumarGross1995,JanikNowakPappZahed1997} to wireless telecommunication~\cite{TseHanly1999,Muller2002,TulinoVerdu2004,SimonMoustakas2004}.

The primary aim of this publication is to advertise the framework of FRV to the audience of quantitative finance (QF) by re--deriving several known results obtained earlier through other (more laborious) methods of Gaussian RMT, as well as solving a few problems for the first time. This illustrates the fluency of the FRV calculus for noisy financial covariances. In particular, we show how an FRV--based back--of--an--envelope calculation leads to a simple equation (\ref{eq:DoublyCorrelatedWishartEquationForMomentsGeneratingFunction}) for a function \smash{$M \equiv M_{\mathbf{c}} ( z )$} (\ref{eq:MomentsGeneratingFunctionDefinition}) that generates all the moments (\ie contains all the spectral information) of a historical estimator $\mathbf{c}$ (\ref{eq:TransformedEstimators}) of the cross--covariance matrix $\mathbf{C}$ (\ref{eq:CovarianceFunctionFactorization}), in the presence of arbitrary ``true'' cross--covariance and auto--covariance matrices $\mathbf{C}$ and $\mathbf{A}$,
\begin{equation}\label{eq:DoublyCorrelatedWishartEquationForMomentsGeneratingFunctionAnnouncement}
z = r M N_{\mathbf{A}} ( r M ) N_{\mathbf{C}} ( M ) .
\end{equation}
Here the $N$'s are certain functions (\ref{eq:BlueFunctionAndNTransformDefinition}), computable once $\mathbf{C}$ and $\mathbf{A}$ are known; and $r \equiv N / T$. (The unrealistic assumption about the Gaussian statistics of the financial assets' returns will be relaxed only in subsequent papers, where generalizations of the current findings, among them (\ref{eq:DoublyCorrelatedWishartEquationForMomentsGeneratingFunctionAnnouncement}), to the L\'{e}vy FRV calculus will be presented; although randomly sampled L\'{e}vy matrices have one or even no finite spectral moments, FRV permits a straightforward analysis of the pertinent moments' generating functions and thus the corresponding spectral distributions.)

This article is organized as follows:
\begin{itemize}
\item The remainder of this section~\ref{s:Introduction} is devoted to discussing the motivations, meaning, and applicability of our results to the financial reality. Our working assumption of Gaussianity, its limitations and possible extensions, have already been touched upon in par.~\ref{sss:TheGaussianApproximation}. In subsec.~\ref{ss:ModelingCrossCorrelations}, various commonly used models for the cross--covariance matrix $\mathbf{C}$ are presented, which may be used as an input for equation (\ref{eq:DoublyCorrelatedWishartEquationForMomentsGeneratingFunctionAnnouncement}). This application is however limited by the appearance of large eigenvalues in the spectrum of $\mathbf{C}$, for which we give a brief justification; it calls for a more refined analysis than currently allowed by FRV. Subsec.~\ref{ss:ModelingAutoCorrelations} deals with auto--covariances $\mathbf{A}$. We show that our method is well--poised to investigate the weak short--ranged auto--correlations observed on financial markets, triggered by non--zero transaction costs and the presence of bonds or interest rates, and included in modern risk evaluation methodologies, but fails at this stage to handle more involved auto--covariances between different assets, such as required for example to explain the Epps effect. We describe also the much more important long--memory auto--correlations between the random volatility of the returns (heteroscedasticity), and introduce some corresponding weighting schemes, mainly the exponentially weighted moving average (EWMA). In subsec.~\ref{ss:HistoricalEstimationOfTheCovarianceMatrices}, we define and discuss several standard historical estimators of the covariance matrices (Pearson, time--lagged, weighted).
\item Section~\ref{s:FreeRandomVariablesAPromisingApproachToTheEstimationOfCovarianceMatrices} is the centerpiece of this work. It commences in subsec.~\ref{ss:TheFreeRandomVariablesCalculusInANutShell} with a crash course in free random variables, with a particular focus on the addition and multiplication algorithms of free random matrices; the latter constitutes the chief tool we exploit. It has a more operational flavor, designed to aid practical applications by the QF community rather than to delve into the mathematics behind the scenes. Subsec.~\ref{ss:TheUncorrelatedWishartEnsembleFromFRV} gives a foretaste of the power of the FRV approach by re--computing in an algebraic way the famous Mar\v{c}enko--Pastur distribution. The salient part comes in subsec.~\ref{ss:TheDoublyCorrelatedWishartEnsembleFromFRV}, where equation (\ref{eq:DoublyCorrelatedWishartEquationForMomentsGeneratingFunctionAnnouncement}), along with its variations, is derived and presented.
\item Section~\ref{s:Examples} contains two examples of how the main formula (\ref{eq:DoublyCorrelatedWishartEquationForMomentsGeneratingFunctionAnnouncement}) can be used to tackle financially relevant problems. Namely, we find equations for the moments' generating functions $M$ of the standard and time--delayed historical estimators in the presence of exponentially decaying temporal auto--covariances and arbitrary cross--covariances (subsec.~\ref{ss:AnExponentiallyDecayingAutoCovariance}), as well as of the EWMA with arbitrary cross--covariances (subsec.~\ref{ss:TheExponentiallyWeightedMovingAverage}). In the simplest case of no underlying cross--covariances, we reinforce our analytical findings with numerical simulations.
\item The article terminates with short conclusions and some possible prospects for the future in section~\ref{s:Conclusions} (more are actually given inside the body of the article), as well as a list of references.
\end{itemize}


\subsection{Modeling Cross--Correlations}
\label{ss:ModelingCrossCorrelations}


\subsubsection{Principal Component Analysis}
\label{sss:PrincipalComponentAnalysis}

In this subsection, we consider a fixed moment in time, $a$, and investigate the cross--correlations between the assets; assuming the decoupling (\ref{eq:CovarianceFunctionFactorization}), their structure $\mathbf{C}$ does not depend on time. Being symmetric and positive--definite, $\mathbf{C}$ can be diagonalized, \smash{$\mathbf{C} \mathbf{v}_{k} = \lambda_{k} \mathbf{v}_{k}$}, with $N$ real and positive eigenvalues (they can be ordered decreasing, \smash{$\lambda_{1} \geq \ldots \geq \lambda_{N} > 0$}), and $N$ orthogonal and normalized eigenvectors. This diagonalization is called a ``principal component analysis'' (PCA), because the knowledge of the eigenvectors of $\mathbf{C}$ allows to linearly transform the $N$ correlated entities \smash{$R_{i a}$} into $N$ uncorrelated ones (referred to as ``principal components'' or ``explicative factors'') \smash{$e_{k a}$}, whose variances are given by the eigenvalues of $\mathbf{C}$,
\begin{equation}\label{eq:PCA}
e_{k a} \equiv \sum_{i = 1}^{N} v_{k , i} R_{i a} , \qquad \textrm{or conversely,} \qquad R_{i a} = \sum_{k = 1}^{N} v_{k , i} e_{k a} , \qquad \textrm{where} \qquad \la e_{k a} e_{l a} \ra = \lambda_{k} \delta_{k l} .
\end{equation}
In other words, the PCA unravels the uncorrelated (not necessarily independent) factors affecting the collection of assets, and orders them w.r.t. their decreasing volatility. Since a factor is a certain mix of the assets (\ie a portfolio), we can restate it yet differently by that the PCA derives a set of uncorrelated investment portfolios, and orders them w.r.t. their decreasing risk.


\subsubsection{Factor Models}
\label{sss:FactorModels}

There have been put forth models of the structure of the covariance matrix (see for example~\cite{BouchaudPotters2003}, section 9.3). They are to reflect the structure of the spectra of its estimators built from historical financial data (see subsec.~\ref{ss:HistoricalEstimationOfTheCovarianceMatrices}), which typically consist of one very large eigenvalue, several smaller but still large ones, and a sea of small eigenvalues, whose distribution can be very accurately fitted with the Mar\v{c}enko--Pastur distribution~\cite{MarcenkoPastur1967} of the eigenvalues of a purely random matrix belonging to the (uncorrelated) Wishart ensemble~\cite{Wishart1928}. As we discuss in more detail in subsec.~\ref{ss:HistoricalEstimationOfTheCovarianceMatrices}, an estimator of the covariance matrix will necessarily contain a lot of measurement noise, and these decade--old results~\cite{LalouxCizeauBouchaudPotters1999,PlerouGopikrishnanRosenowAmaralStanley1999}, pioneering the use of random matrix theory in financial applications, suggest that actually most of the spectrum is purely random, with the exception of the largest eigenvalues ``leaking out'' of the bulk Mar\v{c}enko--Pastur distribution, which do carry some information about the true correlations between the assets. For example, the appearance of one very large eigenvalue \smash{$\lambda_{1}$} (c.a. $25$ times greater than the upper bound of the noise distribution for the S\&P500 data in~\cite{LalouxCizeauBouchaudPotters1999}) has the following meaning: Since \smash{$e_{1}$} fluctuates with such a dominating volatility, the PCA (\ref{eq:PCA}) can be approximated as \smash{$R_{i} \approx v_{1 , i} e_{1}$} (we skip the index $a$ in this paragraph), which means that the dynamics of all the assets is governed practically by just one factor. The corresponding eigenvector has roughly all the components equal, which thus represents a portfolio with approximately the same allocation in all the assets, \ie with no diversification; this portfolio will therefore be strongly correlated with the market index, and is thus called the ``market factor.'' Its presence in the empirical spectrum may be understood for instance in terms of the herding phenomenon (a collective behavior of the investors). Similarly, other large eigenvalues seen in the spectra of historical covariance matrices can be attributed to the clustering of individual companies into industrial sectors (constructed by investigating the relevant eigenvectors), within which the correlations are strong.

Consequently, one may attempt to model the matrix $\mathbf{C}$ in order to reproduce such empirical observations; this program goes under a name of ``factor component analysis'' (FCA), since it aims at describing a large number of cross--correlations between assets in terms of their correlations with a much smaller number of factors. To begin with, one considers a ``one--factor model'' (``market model'')~\cite{Sharpe1964}, where it is assumed that each return \smash{$R_{i}$} is impacted by the market return \smash{$\phi_{0}$} with some strength \smash{$\beta_{i}$} (named the ``market beta'' of this asset), and besides that there is no correlation between assets; namely, it approximates \smash{$R_{i} = \beta_{i} \phi_{0} + e_{i}$}, where \smash{$\phi_{0}$} and the \smash{$e_{i}$}'s (called the ``idiosyncratic noise''; their presence implies that the underlying factors cannot be directly observed as they are marred by random errors) are uncorrelated and have the volatilities \smash{$\Sigma$} and \smash{$\sigma_{i}$} respectively; the model is described by $( 2 N + 1 )$ parameters. The covariance matrix thus reads
\begin{equation}\label{eq:CovarianceMatrixOneFactorModel}
C_{i j} = \Sigma^{2} \beta_{i} \beta_{j} + \sigma_{i}^{2} \delta_{i j} .
\end{equation}
It can be easily diagonalized under an additional simplification that all the \smash{$\sigma_{i}$}'s are equal to some \smash{$\sigma_{0}$}, in which case there is one large ($\propto N$; we generically assume $N$ to be large, see (\ref{eq:ThermodynamicalLimit})) eigenvalue \smash{$\lambda_{1} = \Sigma^{2} \boldsymbol{\beta}^{2} + \sigma_{0}^{2}$}, corresponding to \smash{$\mathbf{v}_{1} \propto \boldsymbol{\beta}$} (the ``market''), and an $( N - 1 )$--degenerate eigenvalue \smash{$\sigma_{0}^{2}$} (if the \smash{$\sigma_{i}$}'s are unequal but of comparable size, there is a large market eigenvalue and a sea of $( N - 1 )$ small ones).

A more refined ``multi--factor model''~\cite{PappPafkaNowakKondor2005,BrinnerConnor2008,Noh2000}, describing increased correlations within industrial sectors, assumes that the idiosyncratic (non--market) parts \smash{$e_{i}$} are exposed to $K$ hidden factors \smash{$\phi_{\alpha}$}, namely \smash{$e_{i} = \sum_{\alpha = 1}^{K} \beta_{i \alpha} \phi_{\alpha} + \epsilon_{i}$}, where all the factors and the new idiosyncratic terms are uncorrelated and have the volatilities \smash{$\Sigma_{\alpha}$} and \smash{$\sigma_{i}$} respectively. In this case,
\begin{equation}\label{eq:CovarianceMatrixMultiFactorModel}
C_{i j} = \Sigma^{2} \beta_{i} \beta_{j} + \sum_{\alpha = 1}^{K} \Sigma_{\alpha}^{2} \beta_{i \alpha} \beta_{j \alpha} + \sigma_{i}^{2} \delta_{i j} .
\end{equation}
To further (quite drastically) simplify this model, we may consider that each asset $i$ is exposed to only one factor $\alpha$, with \smash{$N_{\alpha}$} assets belonging to sector $\alpha$ (\smash{$\sum_{\alpha = 1}^{K} N_{\alpha} = N$}), which translates into \smash{$\beta_{i \alpha} = \beta_{( \alpha^{\prime} \bar{i} ) \alpha} = \delta_{\alpha^{\prime} \alpha} \beta^{( \alpha^{\prime} )}_{\bar{i}}$}, where the index $i$ is split into a double--index \smash{$( \alpha^{\prime} \bar{i} )$}, with \smash{$\alpha^{\prime}$} enumerating sectors and \smash{$\bar{i}$} assets within sector \smash{$\alpha^{\prime}$}. Also we consider the exposures to the market negligible as compared to the industrial ones (\smash{$\beta_{i} = 0$}), and the idiosyncratic volatilities depending only on the sector, \smash{$\sigma_{i} = \sigma_{( \alpha \bar{i} )} = \sigma_{\alpha}$}. Then $\mathbf{C}$ acquires a block--diagonal form, which is easily diagonalized to give $K$ ``large'' eigenvalues \smash{$\lambda_{\alpha} = \Sigma_{\alpha}^{2} \boldsymbol{\beta}^{( \alpha ) 2} + \sigma_{\alpha}^{2}$} plus $K$ ``small'' \smash{$( N_{\alpha} - 1 )$}--degenerate eigenvalues \smash{$\sigma_{\alpha}^{2}$}. It mirrors more properly the existence of several large eigenvalues. The form of the covariance matrix can also be specified along analogous lines in more complex ways, such as in the ``hierarchically nested factor model'' (HNFM)~\cite{TumminelloLilloMantegna2007}.

Any such model can be used as a non--statistical input for equation (\ref{eq:DoublyCorrelatedWishartEquationForMomentsGeneratingFunctionAnnouncement}). In this way, the number of parameters to be estimated (for which the Pearson's chi--square method may be used) is typically greatly reduced, however on the cost of a specification error. Another problem with the above models is that their generic feature is the existence of isolated ``large'' eigenvalues (\ie proportional to the size of the portfolio $N$ and with a microscopic degeneracy, typically singlets), while the FRV tools seem not adequate enough to tackle such situations yet, being restricted to the bulk of the distribution, and thus other methods should be employed (see for example~\cite{BaikBenArousPeche2005}). This is why in this article we refrain from using our main formula (\ref{eq:DoublyCorrelatedWishartEquationForMomentsGeneratingFunctionAnnouncement}) for any nontrivial cross--covariance matrix $\mathbf{C}$, focusing rather on temporal covariances; this obstacle should certainly be dealt with.


\subsection{Modeling Auto--Correlations}
\label{ss:ModelingAutoCorrelations}


\subsubsection{Lagged Correlations Between the Returns}
\label{sss:LaggedCorrelationsBetweenTheReturns}

Let us now discuss which empirical facts concerning temporal correlations can be modeled (and how) within our very simplified framework (\ref{eq:CovarianceFunctionFactorization}). First, it is well--known that the returns are weakly auto--correlated on short time scales: the delayed correlation function (see (\ref{eq:LaggedCovarianceMatrixEstimator}) for its definition) is significantly different from zero (and, for example, negative for stocks, but positive for stock indices) for the time lags less than c.a. $30$ minutes for liquid and free--floating assets (longer on less liquid markets; this decay lag also decreases with time), see~\cite{BouchaudPotters2003}, sections 6.2, 13.1.3. The simplest and most natural model for such a behavior, for a single asset $i$, is an exponential decay,
\begin{equation}\label{eq:ExponentialDecayOfLaggedCorrelationsSingleAsset}
\frac{\la R_{i a} R_{i b} \ra}{\la R_{i a}^{2} \ra} = \ee^{- | b - a | / \tau} ,
\end{equation}
with the characteristic time $\tau$ (given here in the units of $\delta t$) of the order of several minutes.

Let us emphasize that we are talking about \emph{correlation} functions here, \ie normalized by the variance. The variance have completely different, stronger, long--memory temporal dynamics (heteroscedasticity; see par.~\ref{sss:Heteroscedasticity}), allowing forecasts of future volatilities from past data. From this point of view, heteroscedasticity is a ``first--order correction'' to an iid of the returns, while the auto--correlation of the returns (such as (\ref{eq:ExponentialDecayOfLaggedCorrelationsSingleAsset})) is a ``second--order effect.'' Consequently, any long--term forecast of the mean returns seems impossible, and we will focus on forecasting the volatility, used then to assess the short--term risk of a portfolio. Anyway, on short time horizons (such as one business day), the mean return is negligible (say, a small fraction of a percent for stocks) as compared to its volatility (a few percent).

These weak auto--correlations should not however be disregarded. Actually, they may persist on longer time scales, such as days, but to reveal that, one would need to consider much longer (decades) historical time series in order to decrease the estimation noise; there might even be auto--correlations over time spans of a few years, reflecting the existence of economic cycles. If detectable auto--correlations were present for longer lags, they could be used to devise a profitable trading strategy until arbitrage would remove them; this is the efficient market hypothesis. But even such weak and short--memory auto--correlations allow in principle to attain large profits in high--frequency (HF) trading; however, when transaction costs are taken into account, these profits are precisely discounted, and this is one reason that a non--zero decay lag is allowed without contradicting the efficiency of the markets. Another reason is the existence of riskless assets (bonds), which implies that stocks should gain on average the riskless rate of return and additionally a risk premium; moreover, short--term interest rates are not free--floating (they are set by the central banks and are usually quite predictable). For these reasons, the auto--correlations, albeit weak, begin to be included into new risk evaluation methodologies, such as RiskMetrics~2006~\cite{RiskMetrics2006}, which even attempts to predict the mean returns (and thus risks) for long--time horizons, up to one year. Our approximation (\ref{eq:CovarianceFunctionFactorization}) should be able to handle effects like (\ref{eq:ExponentialDecayOfLaggedCorrelationsSingleAsset}), and in subsec.~\ref{ss:AnExponentiallyDecayingAutoCovariance} we indeed show an application of equation (\ref{eq:DoublyCorrelatedWishartEquationForMomentsGeneratingFunctionAnnouncement}) to a model with an exponentially decaying auto--covariance matrix $\mathbf{A}$.

However, this phenomenon of non--zero lagged correlations should be extended from a single asset to \emph{multiple} variables: the returns of different assets are correlated between different time moments. This is crucial, for example, for gaining insight into the dynamics of the cross--correlations as we progress from the HF time scale to longer time scales. The strength of the equal--time cross--correlation between assets $i \neq j$ was long ago observed to grow with increasing $\delta t$ (when moving from higher to lower sampling frequencies, the equal--time inter--asset correlations rapidly rise on the scales of several minutes, to saturate on the scales of days), which is called the ``Epps effect''~\cite{Epps1979}. Not only this, but the entire structure of cross--correlations (depicted through the maximum spanning tree of the market~\cite{BonannoVandewalleMantegna2000}) evolves as an embryo which expands and differentiates as $\delta t$ enlarges. This change of strength and structure of cross--correlations is, for instance, critical for the possibility of increasing the sampling frequency in order to obtain longer time series, and as a result, less noisy historical estimators (see par.~\ref{sss:EstimatorsOfCrossCovariances}): one cannot probe too deep into the HF regime (far from the saturation level of the Epps curve) since then the entire tree of cross--correlations looks totally different. This is yet another reason for developing noise--cleaning procedures, such as the one advocated in this paper. In order to explain the Epps effect, a causal relation must be present between the time evolution of the return of asset $i$ at a certain time and the returns of all the other assets $j \neq i$ at all the previous moments. For example, in~\cite{TothTothKertesz2007,TothKertesz2007}, the equal--time cross--correlations are expressed through delayed cross--correlations over shorter time scales, and the latter are modeled by a direct analog of the exponential decay (\ref{eq:ExponentialDecayOfLaggedCorrelationsSingleAsset}), just with separate $i$ and $j$,
\begin{equation}\label{eq:ExponentialDecayOfLaggedCorrelationsDifferentAssets}
\frac{\la R_{i a} R_{j b} \ra}{\la R_{i a} R_{j a} \ra} = \ee^{- | b - a | / \tau} ;
\end{equation}
this eventually proves to provide an analytical shape of the Epps curve which is remarkably close to the experimental one. Another, more complex model of linear causal influence is presented in~\cite{PottersBouchaudLaloux2005},
\begin{equation}\label{eq:LinearCausalRegressionModel}
r_{i} ( t ) = e_{i} ( t ) + \sum_{j = 1}^{N} \int_{- \infty}^{+ \infty} \dd t^{\prime} K_{i j} \left( t - t^{\prime} \right) r_{j} \left( t^{\prime} \right) ,
\end{equation}
where \smash{$e_{i} ( t )$} is an idiosyncratic part, and \smash{$K_{i j} ( t - t^{\prime} )$} is called the ``influence kernel.'' However, these models cannot be captured by our current simple framework (\ref{eq:CovarianceFunctionFactorization}), (\ref{eq:DoublyCorrelatedWishartEquationForMomentsGeneratingFunctionAnnouncement}), and it certainly is an interesting challenge to extend the FRV approach so that it could be helpful in an analytical treatment of such more involved correlations, responsible among other things for the Epps effect.


\subsubsection{Heteroscedasticity}
\label{sss:Heteroscedasticity}

A well--established ``stylized fact'' observed in all financial time series is that the (say, daily) volatility of an asset's return depends on time, displaying a ``long memory,'' namely that periods of high or low volatility tend to persist over time; this property is known as ``heteroscedasticity,'' ``volatility clustering,'' or ``intermittence'' (by analogy with turbulent flows of fluids, where a similar phenomenon of persistent intertwined periods of laminar and turbulent behavior occurs). A standard approach to describe mathematically this experimental fact is to suppose that not only is the demeaned and normalized return \smash{$\epsilon_{i a}$} (the ``residual'') a random variable, but so is the volatility \smash{$\sigma_{i a}$},
\begin{equation}\label{eq:StochasticVolatilityModel}
R_{i a} = \sigma_{i a} \epsilon_{i a} .
\end{equation}
These two sources of randomness are to a great extent inseparable, and it becomes a matter of choice how to model them in order to jointly arrive at results which comply with empirical data; for example, a Student distribution for the return can originate from an inverse--gamma randomness of the variance superimposed on a Gaussian iid of the residuals. Such a very general depiction (\ref{eq:StochasticVolatilityModel}), with the \smash{$\epsilon_{i a}$}'s assumed to be iid with zero mean and unit volatility and the \smash{$\sigma_{i a}$}'s some random variables possibly correlated with each other and also with the residuals, is named a ``stochastic volatility model''; see~\cite{BouchaudPotters2003}, section 7.

Indeed, the volatilities at different time moments are correlated. These lagged correlations are not strong (several percent, depending on the volatility proxy used and the time lag chosen), but stretch over much longer periods than the lagged correlations of the residuals, discussed in par.~\ref{sss:LaggedCorrelationsBetweenTheReturns}: their slow decay (long--memory) can be modeled well by a power law, \smash{$1 / | b - a |^{\nu}$}, where a fit of the exponent $\nu$ typically lies in the range $0.2 \div 0.4$ (depending on the domain of the time lags considered). In other words, the temporal dynamics of the volatility is a multi--scale phenomenon. Moreover, the probability distribution of (an estimator, such as the ``high--frequency proxy,'' being the daily average of HF returns, of) the volatility may be approximated by an inverse--gamma or a log--normal shape.

A basic idea, founded upon the presence of the long--memory lagged volatility correlations, is to regard the volatility as undergoing a certain \emph{stochastic process}. A convenient feature of this approach is its consistency: the volatility process should be constructed from historical time series (in particular, it should reflect a power--law--like decay of the lagged correlations), and thus obtained parameters are then used to make forecasts (through evaluating conditional averages) for the value of the volatility over some future time horizon $\Delta t$. In other words, even long--time horizons become available for meaningful risk assessment; although of course for long $\Delta t$ the deficiency of past data excludes a possibility of backtesting of these forecasts. There exists a plethora of propositions for volatility processes. One selecting requirement is actually computational accessibility of forecasting, which practically reduces the possible choices to quadratic processes only, \ie where the variance depends linearly on the past squared returns, see below. This still yields a very broad class of processes, falling under the name of ``auto--regressive conditional heteroscedasticity'' (ARCH).

A standard textbook example, reflecting to some extent the behavior of financial time series, is the GARCH$( 1 , 1 )$ model~\cite{Engle1982,EngleBollerslev1986,Bollerslev1986},
\begin{equation}\label{eq:GARCHOneOne}
\sigma_{a}^{2} = w_{\infty} \sigma_{\mathrm{mean}}^{2} + \left( 1 - w_{\infty} \right) \sigma_{\mathrm{hist.} , a}^{2} , \qquad \sigma_{\mathrm{hist.} , a}^{2} = \alpha \sigma_{\mathrm{hist.} , a - 1}^{2} + ( 1 - \alpha ) R_{a - 1}^{2}
\end{equation}
(the asset index $i$ is skipped here and in the remainder of this paragraph). In the above, \smash{$\sigma_{\mathrm{mean}}^{2}$} is the unconditional mean variance, representing the average long--run value of the variance, and \smash{$\sigma_{\mathrm{hist.} , a}^{2}$} is a ``historical (auto--regressive) variance,'' depending linearly on both itself and the realized squared return at the preceding time moment (the daily frequency is typically used). The constant \smash{$w_{\infty}$} is a ``coupling,'' while $\alpha \in [ 0 , 1 ]$ may be thought of as measuring the responsiveness of the variance to the recent realized variance \smash{$R_{a - 1}^{2}$}: $\alpha$ close to $1$ means that the variance responds quite slowly to the news. (We may also write (\ref{eq:GARCHOneOne}) differently, \smash{$\sigma_{a}^{2} = \sigma_{\mathrm{mean}}^{2} + g_{1} ( \sigma_{a - 1}^{2} - \sigma_{\mathrm{mean}}^{2} ) + g_{2} ( R_{a - 1}^{2} - \sigma_{a - 1}^{2} )$}, where \smash{$g_{1} \equiv 1 - w_{\infty} ( 1 - \alpha )$} and \smash{$g_{2} \equiv ( 1 - w_{\infty} ) ( 1 - \alpha )$}, in order to see that the process tries to revert the volatility to its mean value with strength \smash{$g_{1}$}, and also incorporates a feedback of the difference between the realized squared return \smash{$R_{a - 1}^{2}$} and its mean value \smash{$\sigma_{a - 1}^{2}$} on the next--day value of the variance, to which effect a magnitude \smash{$g_{2}$} is given.)

There are two problems with this traditional model. First, it is an ``affine'' (``mean--reverting'') process, \ie containing the additive term \smash{$w_{\infty} \sigma_{\mathrm{mean}}^{2}$}, whose meaning is that in the long run, the average volatility will converge to the value \smash{$\sigma_{\mathrm{mean}}$} regardless of the initial conditions. To every affine process, there exists a corresponding ``linear'' (``integrated'') process, which simply removes the unconditional expectation by setting \smash{$w_{\infty} = 0$}, in which case the long--term mean volatility depends on the initial conditions or may even not converge; for example, (\ref{eq:GARCHOneOne}) will yield a model called I--GARCH$( 1 )$. Only the integrated processes can be successfully harnessed for risk forecasts. The reason is that an integrated model is described by two parameters less than its mean--reverting counterpart (namely, \smash{$w_{\infty}$} and \smash{$\sigma_{\mathrm{mean}}$}), and the latter of these is strongly time series dependent; in other words, for a portfolio of $N$ assets, an affine model would contribute a large number $N$ of mean unconditional volatilities for estimation, which would produce a huge measurement error. In I--GARCH$( 1 )$, on the other hand, there is only one parameter $\alpha$ accounting for all the assets that requires estimation.

More importantly, both GARCH$( 1 , 1 )$ and I--GARCH$( 1 )$ (let us henceforth focus on the integrated versions only) fail to reproduce the observed power--law decay of the time--lagged volatility correlations, leading instead (the calculation is doable analytically) to an exponential decay, with the characteristic time (in the units of $\delta t = \textrm{one day}$) \smash{$\tau = - 1 / \log \alpha$},
\begin{equation}\label{eq:GARCHOneOneExponentialDecay}
\la \sigma_{a}^{2} \sigma_{b}^{2} \ra - \la \sigma_{a}^{2} \ra \la \sigma_{b}^{2} \ra \propto \ee^{- | b - a | / \tau} .
\end{equation}
This may be seen in yet another way by unwinding the second part of (\ref{eq:GARCHOneOne}), thus casting the variance as a linear function of the past squared returns,
\begin{equation}\label{eq:EWMAWeights}
\sigma_{a}^{2} = \frac{1 - \alpha}{1 - \alpha^{T}} \sum_{b = 1}^{T} \alpha^{b - 1} R_{a - b}^{2} ,
\end{equation}
where a necessary cutoff $T$ is introduced. The ``weights'' with which the past squared returns impact the today's variance, scale exponentially as we move backward in time (\smash{$\propto \alpha^{b - 1}$}); this short--memory scheme is called the ``exponentially weighted moving average'' (EWMA), see for example~\cite{Hull2008}, chapter 21, and~\cite{PafkaPottersKondor2004,Svensson2007}. Despite this evident shortcoming, the I--GARCH$( 1 )$ (EWMA) volatility process has very successfully transpired into the every--day practice of many financial institution by being woven into the commonly accepted risk evaluation methodology, RiskMetrics~1994~\cite{RiskMetrics1996,MinaXiao2001}. It was probably due to its simplicity, as it is described by just one parameter $\alpha$ shared by a wide range of securities (its value found to yield forecasts which come closest to the realized variance is $\alpha = 0.94$, \ie $\tau = 16.2 \textrm{ business days}$), which is critical since the methodology is to be applied to a great many time series; and moreover, it utilizes only the one previous--day observation to update the volatility (so little data needs to be stored).

Let us briefly mention that the framework of integrated models can be extended to accommodate for the long--memory correlations~\cite{Zumbach2004,Zumbach2009}, culminating in the fresh RiskMetrics~2006~\cite{RiskMetrics2006}. Such processes are still quadratic,
\begin{equation}\label{eq:GeneralWeights}
\sigma_{a}^{2} = \sum_{b = 1}^{T} w_{b} R_{a - b}^{2} ,
\end{equation}
where the weights \smash{$w_{b}$} are positive and obey the ``sum rule,'' \smash{$\sum_{b = 1}^{T} w_{b} = 1$}. For example, RiskMetrics~2006 argues that a logarithmic decay consistently proves to be an even better fit to financial data than a power law,
\begin{equation}\label{eq:RiskMetrics2006Weights}
w_{b} \propto 1 - \frac{\log ( b \delta t )}{\log \tau_{0}} ,
\end{equation}
where again one parameter \smash{$\tau_{0} \sim 3 \div 6 \textrm{ years}$} is enough to capture the long memory of diverse time series. (For small time lags, the power--law and log--decay are very similar, with their parameters related approximately through \smash{$\nu = 1 / \log ( \tau_{0} / \delta t )$}. But for longer lags, say beyond one month, the logarithmic fit visibly stands out in quality.)

Our FRV technique is not yet suited for handling ARCH models like discussed above. However, the FRV calculus does bring considerable simplification into working with historical estimators of cross--covariance matrices which incorporate weighting schemes (\ref{eq:GeneralWeights}), such as the EWMA (\ref{eq:EWMAWeights}) or log--decay (\ref{eq:RiskMetrics2006Weights}). They will be defined in par.~\ref{sss:EstimatorsWithWeightingSchemes}, and the case of the EWMA (with all its defects and advantages just highlighted) will be addressed in subsec.~\ref{ss:TheExponentiallyWeightedMovingAverage}.


\subsection{Historical Estimation of the Covariance Matrices}
\label{ss:HistoricalEstimationOfTheCovarianceMatrices}


\subsubsection{Estimators of Equal--Time Cross--Covariances}
\label{sss:EstimatorsOfCrossCovariances}

A fundamental problem is how to reliably estimate the covariance matrices from the available historical data~\cite{SilversteinBai1995,SilversteinBai2006}. One obstacle lies in the finiteness of the time series, due to which any estimator will contain an amount of measurement noise. Let us focus for definiteness on estimating $\mathbf{C}$: since there are $N ( N + 1 ) / 2$ independent entries in $\mathbf{C}$, and we have at our disposal $N$ time series of length $T$ each (collected in a historical realization $\mathbf{R}$; we will not distinguish in notation between random variables and their actual realizations), thus the level of the estimation noise may be quantified by the ``rectangularity ratio''
\begin{equation}\label{eq:NoiseToSignalRatio}
r \equiv \frac{N}{T} .
\end{equation}
If $r \to 0$ (thanks to $T \to \infty$ with fixed $N$, which is a limit commonly used in statistics), any empirical covariance matrix should approach the exact one (\ie it should be asymptotically unbiased). However, $r$ close to zero is usually far from financial reality, in which both $T$ and $N$ are large and of comparable size; for example, one may have daily data from several years (each of about $260$ business days) and may want to consider a major bank's portfolio consisting of several hundred of even thousands of assets; hence, the relevant regime is rather the ``thermodynamical limit,''
\begin{equation}\label{eq:ThermodynamicalLimit}
N \to \infty , \qquad T \to \infty , \qquad \textrm{such that} \qquad r = \textrm{fixed} .
\end{equation}
Therefore, any estimator will be (seriously, for realistic values of $r$) dressed with the measurement noise, and it is of paramount importance (from the point of view of risk management, for example) to devise methods which detect in the noised estimators information about the true covariances (``cleaning'' of the measurement errors); these de--noised estimators can then serve for practical purposes (such as evaluating the risk of a portfolio). (Remark that (\ref{eq:ThermodynamicalLimit}) is also precisely the limit in which the standard techniques of random matrix theory are applicable to the random matrix $\mathbf{R}$; to be used below.)

The matrices $\mathbf{C}$ and $\mathbf{A}$ can be estimated from the past time series $\mathbf{R}$ by, for example,
\begin{equation}\label{eq:EstimatorsDefinitions}
\mathbf{c} \equiv \frac{1}{T} \mathbf{R} \mathbf{R}^{\TT} , \qquad \mathbf{a} \equiv \frac{1}{N} \mathbf{R}^{\TT} \mathbf{R} ,
\end{equation}
which may be called their Pearson estimators (the usual prefactors $1 / ( T - 1 )$ and $1 / ( N - 1 )$ are replaced in the above by $1 / T$ and $1 / N$, respectively, since we can approximately disregard the average value of the returns in comparison with their volatilities over the considered time horizons; see par.~\ref{sss:LaggedCorrelationsBetweenTheReturns}). For any probability distribution of the returns such that (\ref{eq:CovarianceFunctionFactorization}) holds with finite $\mathbf{C}$ and $\mathbf{A}$, it is easily checked that the Pearson estimators (\ref{eq:EstimatorsDefinitions}) are, up to rescalings, unbiased,
\begin{equation}\label{eq:EstimatorsBias}
\la \mathbf{c} \ra = M_{\mathbf{A} , 1} \mathbf{C} , \qquad \la \mathbf{a} \ra = M_{\mathbf{C} , 1} \mathbf{A} ,
\end{equation}
where \smash{$M_{\mathbf{C} , 1} \equiv \frac{1}{N} \Tr \mathbf{C}$} and \smash{$M_{\mathbf{A} , 1} \equiv \frac{1}{T} \Tr \mathbf{A}$} are the first moments of the matrices $\mathbf{C}$ and $\mathbf{A}$, see below (\ref{eq:MomentsExpansionOfTheGreenFunction}). For the correlated Gaussian distribution (\ref{eq:CorrelatedGaussian}), the Pearson estimators are proportional to the respective maximum likelihood estimators.

It is clear that \smash{$c_{i j} = \frac{1}{T} \sum_{a = 1}^{T} R_{i a} R_{j a}$} represents the equal--time covariance between assets $i$ and $j$ averaged over time; similarly, \smash{$a_{a b} = \frac{1}{N} \sum_{i = 1}^{N} R_{i a} R_{i b}$} shows how the measurements at moments $a$ and $b$ are correlated on average for all the assets. These two seemingly very different quantities are in fact very closely related: since \smash{$\mathbf{R} \mathbf{R}^{\TT}$} and \smash{$\mathbf{R}^{\TT} \mathbf{R}$} have the same non--zero eigenvalues (the larger one has additionally $| T - N |$ zero modes), thus $\mathbf{c}$ and $\mathbf{a}$ have also identical non--zero eigenvalues up to the factor of $r$ (the latter are $1 / r$ times the former). In other words, the information content of $\mathbf{c}$ and $\mathbf{a}$ is equivalent, describing the structure of equal--time correlations between the assets; we will thus abandon $\mathbf{a}$ henceforth. We will state this point in more quantitative terms (\ref{eq:DualityForTheMomentsGeneratingFunctionAndGreenFunction}) in par.~\ref{sss:TheDuality}.

As is well--known, it is possible to describe the $N \times T$ correlated Gaussian random variables $\mathbf{R}$ in terms of $N \times T$ uncorrelated Gaussian variables \smash{$\widetilde{\mathbf{R}}$}; this is achieved through the change \smash{$\mathbf{R} = \sqrt{\mathbf{C}} \widetilde{\mathbf{R}} \sqrt{\mathbf{A}}$} (the covariance matrices are symmetric and positive--definite, therefore their square roots exist), which transforms the correlated Gaussian measure (\ref{eq:CorrelatedGaussian}) into the uncorrelated one,
\begin{equation}\label{eq:UncorrelatedGaussian}
P_{\mathrm{G.}} ( \widetilde{\mathbf{R}} ) \DD \widetilde{\mathbf{R}} = \frac{1}{\mathcal{N}_{\mathrm{G.}}} \exp \left( - \frac{1}{2} \Tr \widetilde{\mathbf{R}}^{\TT} \widetilde{\mathbf{R}} \right) \DD \widetilde{\mathbf{R}} = \frac{1}{\mathcal{N}_{\mathrm{G.}}} \exp \left( - \frac{1}{2} \sum_{i = 1}^{N} \sum_{a = 1}^{T} \widetilde{R}_{i a}^{2} \right) \DD \widetilde{\mathbf{R}} ,
\end{equation}
with \smash{$\mathcal{N}_{\mathrm{G.}} = ( 2 \pi )^{N T / 2}$}, and \smash{$\langle \ldots \rangle_{\mathrm{G.}}$} denoting the expectation map w.r.t. this probability distribution. Correspondingly, the estimator $\mathbf{c}$ becomes in the new language more involved,
\begin{equation}\label{eq:TransformedEstimators}
\mathbf{c} = \frac{1}{T} \sqrt{\mathbf{C}} \widetilde{\mathbf{R}} \mathbf{A} \widetilde{\mathbf{R}}^{\TT} \sqrt{\mathbf{C}} .
\end{equation}
With \smash{$\widetilde{\mathbf{R}}$} a random matrix drawn from the distribution (\ref{eq:UncorrelatedGaussian}), the estimator $\mathbf{c}$ (\ref{eq:TransformedEstimators}) is called a ``doubly correlated Wishart'' random matrix.


\subsubsection{Estimators of Time--Delayed Cross--Covariances}
\label{sss:EstimatorsOfTimeDelayedCrossCovariances}

It is of course desirable to find an estimator of temporal correlations, \ie correlations between two assets at two different moments in time. It is commonly done through the ``lagged covariance matrix estimator,'' which represents non--equal--time (with a time lag $d$, an integer divisor of the total time series length $T$, $t \equiv T / d = 2 , 3 , \ldots$) covariance between assets $i$ and $j$ averaged over time,
\begin{equation}\label{eq:LaggedCovarianceMatrixEstimator}
c_{i j}^{( d )} \equiv \frac{1}{T} \sum_{a = 1}^{T - d} R_{i a} R_{j , a + d} , \qquad \textrm{\ie} \qquad \mathbf{c}^{( d )} = \frac{1}{T} \mathbf{R} \mathbf{D}^{( d )} \mathbf{R}^{\TT} , \qquad \textrm{where} \qquad D_{a b}^{( d )} \equiv \delta_{a + d , b} .
\end{equation}
This matrix is non--symmetric, and it will be very interesting to develop a method to deal with it (see~\cite{ThurnerBiely2007} for a solution, based on the circular symmetry of the problem and the Gaussian approximation, in the simplest case of \smash{$\mathbf{C} = \mathbf{1}_{N}$} and \smash{$\mathbf{A} = \mathbf{1}_{T}$}; here \smash{$\mathbf{1}_{K}$} denotes the unit $K \times K$ matrix). In the present paper, however, we will not attempt this more challenging task, leaving it for future work, but only resort to the simplification~\cite{MayyaAmritkar2006} of considering a symmetrized version of (\ref{eq:LaggedCovarianceMatrixEstimator}),
\begin{equation}\label{eq:SymmetrizedLaggedCovarianceMatrixEstimator}
\mathbf{c}^{\mathrm{sym.} ( d )} \equiv \frac{1}{T} \mathbf{R} \mathbf{D}^{\mathrm{sym.} ( d )} \mathbf{R}^{\TT} , \qquad \textrm{where} \qquad D^{\mathrm{sym.} ( d )}_{a b} \equiv \frac{1}{2} \left( \delta_{a + d , b} + \delta_{a - d , b} \right) ,
\end{equation}
which becomes symmetric, and therefore tractable within our present approach, but still carries some information about delayed correlations between assets. In terms of the uncorrelated Gaussian variables (\ref{eq:UncorrelatedGaussian}) it reads
\begin{equation}\label{eq:TransformedSymmetrizedLaggedCovarianceMatrixEstimator}
\mathbf{c}^{\mathrm{sym.} ( d )} = \frac{1}{T} \sqrt{\mathbf{C}} \widetilde{\mathbf{R}} \sqrt{\mathbf{A}} \mathbf{D}^{\mathrm{sym.} ( d )} \sqrt{\mathbf{A}} \widetilde{\mathbf{R}}^{\TT} \sqrt{\mathbf{C}} .
\end{equation}
This is also a doubly correlated Wishart random matrix, akin to $\mathbf{c}$, albeit with a modified underlying auto--covariance matrix, \smash{$\mathbf{A} \to \sqrt{\mathbf{A}} \mathbf{D}^{\mathrm{sym.} ( d )} \sqrt{\mathbf{A}}$}.


\subsubsection{Estimators with Weighting Schemes}
\label{sss:EstimatorsWithWeightingSchemes}

The standard Pearson estimator \smash{$c_{i j}$} of the cross--covariance between assets $i$ and $j$ (\ref{eq:EstimatorsDefinitions}) is defined as the average of the realized cross--covariances \smash{$R_{i a} R_{j a}$} over the past time moments $a$. In this way, all these past values of the realized cross--covariance have an equal impact on the estimator of the today's cross--covariance. However, when discussing an analogous problem for the estimates of the variance in par.~\ref{sss:Heteroscedasticity}, we discovered that the phenomenon of heteroscedasticity, modeled by some quadratic ARCH stochastic process (\ref{eq:GeneralWeights}), implies the presence in financial time series of a long memory, described by the weights \smash{$w_{a}$} (of a power--law or logarithmic decay, but frequently used as well is an exponential decay, \ie the EWMA). In other words, the older the realized variance \smash{$R_{i a}^{2}$}, the more obsolete it is, \ie the more suppressed its contribution to the estimator of the today's variance is, as given by the weight \smash{$w_{a}$}. (Here we will adopt a convention that in our time series, enumerated by $a = 1 , \ldots , T$, the most recent observation is $a = 1$, and moving backward in time as $a$ increases.) Now, it is a common practice to set up the updating schemes for the cross--covariances by simply mimicking the schemes for the variances; see~\cite{Hull2008}, compare also the discussion in~\cite{Zumbach2009}. Therefore, we will consider the following general class of ``weighted estimators'' of the cross--covariances,
\begin{equation}\label{eq:WeightedEstimatorDefinitions}
c^{\mathrm{weight}}_{i j} \equiv \sum_{a = 1}^{T} w_{a} R_{i a} R_{j a} , \qquad \textrm{\ie} \qquad \mathbf{c}^{\mathrm{weight}} = \frac{1}{T} \mathbf{R} \mathbf{W} \mathbf{R}^{\TT} , \qquad \textrm{where} \qquad \mathbf{W} \equiv T \diag \left( w_{1} , \ldots , w_{T} \right) .
\end{equation}

Again, it is convenient to convert the correlated Gaussian random variables $\mathbf{R}$ into the uncorrelated ones \smash{$\widetilde{\mathbf{R}}$}, which leads to an expression analogous to (\ref{eq:TransformedSymmetrizedLaggedCovarianceMatrixEstimator}),
\begin{equation}\label{eq:TransformedWeightedEstimator}
\mathbf{c}^{\mathrm{weight}} = \frac{1}{T} \sqrt{\mathbf{C}} \widetilde{\mathbf{R}} \sqrt{\mathbf{A}} \mathbf{W} \sqrt{\mathbf{A}} \widetilde{\mathbf{R}}^{\TT} \sqrt{\mathbf{C}} ,
\end{equation}
which is the doubly correlated Wishart ensemble with the underlying covariance matrices $\mathbf{C}$ and \smash{$\sqrt{\mathbf{A}} \mathbf{W} \sqrt{\mathbf{A}}$}. In this way, also the weighted estimators have been grasped by our general framework.


\section{Free Random Variables: A Promising Approach to the Estimation of Covariance Matrices}
\label{s:FreeRandomVariablesAPromisingApproachToTheEstimationOfCovarianceMatrices}


\subsection{The Free Random Variables Calculus in a Nut--Shell}
\label{ss:TheFreeRandomVariablesCalculusInANutShell}


\subsubsection{The Basic Notions of Random Matrix Theory}
\label{sss:TheBasicNotionsOfRandomMatrixTheory}

When studying (see for example~\cite{Mehta2004,Eynard2000}) a real symmetric (or complex Hermitian) $K \times K$ random matrix $\mathbf{H}$, drawn from some probability distribution $P ( \mathbf{H} )$, perhaps a most natural question is about the probability distribution of its (real) eigenvalues \smash{$\lambda_{1} , \ldots , \lambda_{K}$}, which is quantified by the ``mean spectral density,''
\begin{equation}\label{eq:MeanSpectralDensityDefinition}
\rho_{\mathbf{H}} ( \lambda ) \equiv \frac{1}{K} \sum_{i = 1}^{K} \la \delta \left( \lambda - \lambda_{i} \right) \ra = \frac{1}{K} \la \Tr \left( \lambda \mathbf{1}_{K} - \mathbf{H} \right) \ra ,
\end{equation}
where $\delta ( \lambda )$ is the real Dirac delta function, the expectation map $\langle \ldots \rangle$ is performed w.r.t. $P ( \mathbf{H} )$, and we recall that \smash{$\mathbf{1}_{K}$} denotes the unit $K \times K$ matrix.

This statistical information about the spectrum is equivalently encoded in the ``Green's function'' (also called ``resolvent,'' ``Cauchy transform'' or ``Stieltjes transform''), which is a complex function of a complex variable $z$,
\begin{equation}\label{eq:GreenFunctionDefinition}
G_{\mathbf{H}} ( z ) \equiv \frac{1}{K} \sum_{i = 1}^{K} \la \frac{1}{z - \lambda_{i}} \ra = \frac{1}{K} \la \Tr \frac{1}{z \mathbf{1}_{K} - \mathbf{H}} \ra = \int_{\mathrm{cuts}} \dd \lambda \rho_{\mathbf{H}} ( \lambda ) \frac{1}{z - \lambda} .
\end{equation}
For finite $K$, this is a meromorphic function, with the poles at the \smash{$\lambda_{i}$}'s on the real axis. On the other hand, in the usually considered limit of an infinitely large random matrix ($K \to \infty$), the mean eigenvalues tend to merge into continuous intervals (``cuts''; they can be infinite or finite, connected or not), and the Green's function becomes holomorphic everywhere on the complex plane except the cuts on the real line. As such, it can typically be expanded into a power series around $z \to \infty$,
\begin{equation}\label{eq:MomentsExpansionOfTheGreenFunction}
G_{\mathbf{H}} ( z ) = \sum_{n \geq 0} \frac{M_{\mathbf{H} , n}}{z^{n + 1}} , \qquad M_{\mathbf{H} , n} \equiv \frac{1}{K} \la \Tr \mathbf{H}^{n} \ra = \int_{\mathrm{cuts}} \dd \lambda \rho_{\mathbf{H}} ( \lambda ) \lambda^{n} ,
\end{equation}
where the coefficients are called the ``moments'' of $\mathbf{H}$. In particular, in the strict limit $z \to \infty$, it must obey
\begin{equation}\label{eq:GreenFunctionAtInfinity}
G_{\mathbf{H}} ( z ) \to \frac{1}{z} , \qquad \textrm{for} \qquad z \to \infty .
\end{equation}
The above expansion (\ref{eq:MomentsExpansionOfTheGreenFunction}) suggests working with an alternative object to the Green's function, namely the ``generating function of the moments'' (or the ``$M$--transform''), simply related to the former,
\begin{equation}\label{eq:MomentsGeneratingFunctionDefinition}
M_{\mathbf{H}} ( z ) \equiv z G_{\mathbf{H}} ( z ) - 1 = \sum_{n \geq 1} \frac{M_{\mathbf{H} , n}}{z^{n}} .
\end{equation}
We will be using both, depending on convenience, but chiefly (\ref{eq:MomentsGeneratingFunctionDefinition}). However, we stress that even if the moments do not exist, and thus the expansions (\ref{eq:MomentsExpansionOfTheGreenFunction}), (\ref{eq:MomentsGeneratingFunctionDefinition}) are not valid, the knowledge of the analytical structure of the Green's function (\ref{eq:GreenFunctionDefinition}) is sufficient to extract the statistical spectral properties of the random matrix.

Namely, once the Green's function has been derived, the corresponding mean spectral density is found by using the Sokhotsky's formula, \smash{$\lim_{\epsilon \to 0^{+}} 1 / ( \lambda + \ii \epsilon ) = \mathrm{pv} ( 1 / \lambda ) - \ii \pi \delta ( \lambda )$}, which yields
\begin{equation}\label{eq:SokhotskyFormula}
\rho_{\mathbf{H}} ( \lambda ) = - \frac{1}{\pi} \lim_{\epsilon \to 0^{+}} \im G_{\mathbf{H}} ( \lambda + \ii \epsilon ) .
\end{equation}
In other words, the density is inferred from the behavior of the Green's function in the imaginary vicinity of the eigenvalues' cuts on the real axis.

Finally, let us introduce the functional inverses of the Green's function and the moments' generating function,
\begin{equation}\label{eq:BlueFunctionAndNTransformDefinition}
G_{\mathbf{H}} \left( B_{\mathbf{H}} ( z ) \right) = B_{\mathbf{H}} \left( G_{\mathbf{H}} ( z ) \right) = z , \qquad M_{\mathbf{H}} \left( N_{\mathbf{H}} ( z ) \right) = N_{\mathbf{H}} \left( M_{\mathbf{H}} ( z ) \right) = z .
\end{equation}
The former has somewhat fancifully been named~\cite{Zee1996} the ``Blue's function'' (known also under other names in literature), while the latter will more conservatively be called the ``$N$--transform.'' These two functions are fundamental objects within the FRV approach, see below. Additionally, the Blue's function can be expanded into a power series around $z = 0$: it must start from a singular term $1 / z$ due to (\ref{eq:GreenFunctionAtInfinity}) plus a regular expansion,
\begin{equation}\label{eq:FreeCumulantsDefinition}
B_{\mathbf{H}} ( z ) = \frac{1}{z} + \sum_{n \geq 0} K_{\mathbf{H} , n + 1} z^{n} ,
\end{equation}
where the coefficients, for the reason explained below, are referred to as ``free cumulants.'' (Let us mention that there is another commonly exploited object equivalent to the Blue's function, which subtracts the singular term from the above expansion, and is named the ``$R$--transform,'' \smash{$R_{\mathbf{H}} ( z ) \equiv B_{\mathbf{H}} ( z ) - 1 / z$}. We will however adhere to using the Blue's function.)


\subsubsection{The Basic Notions of the Free Random Variables Calculus}
\label{sss:TheBasicNotionsOfTheFreeRandomVariablesCalculus}

Let us now detail the key features of the so--called ``free random variables'' (FRV) calculus, presented parallel to the corresponding notions in the standard probability calculus.

An important problem in classical probability~\cite{Feller} is to find the probability density function (PDF) of the sum of two random variables, \smash{$x_{1} + x_{2}$}, provided they are independent, and we are given their separate PDFs, \smash{$P_{x_{1}}$} and \smash{$P_{x_{2}}$}. This is readily solved by applying the Newton's formula to the moments of the sum, \smash{$M_{x_{1} + x_{2} , n} = \langle ( x_{1} + x_{2} )^{n} \rangle = \sum_{k = 0}^{n} \binom{n}{k} M_{x_{1} , k} M_{x_{2} , n - k}$}. The moments are conveniently encoded in terms of the ``characteristic function,'' which is a Fourier transform of the PDF,
\begin{equation}\label{eq:CharacteristicFunctionDefinition}
g_{x} ( z ) \equiv \sum_{n \geq 0} \frac{M_{x , n}}{n!} z^{n} = \langle e^{z x} \rangle .
\end{equation}
(Here $z$ must be a purely imaginary number on account of convergence of the sum over $n$, but we will not explicitly print this for the sake of future reference.) The above addition rule for the moments can be re--stated as that the characteristic function is multiplicative under the addition of independent random variables. In other words, its logarithm,
\begin{equation}\label{eq:LogarithmOfTheCharacteristicFunctionDefinition}
r_{x} ( z ) \equiv \log g_{x} ( z ) ,
\end{equation}
is additive,
\begin{equation}\label{eq:TheClassicalAdditionLaw}
r_{x_{1} + x_{2}} ( z ) = r_{x_{1}} ( z ) + r_{x_{2}} ( z ) , \qquad \textrm{for independent \smash{$x_{1}$}, \smash{$x_{2}$}.}
\end{equation}
This may be named the ``classical addition law''; it shows that the addition problem for classical independent random variables is solved by (i) forming the Fourier transforms of the PDFs \smash{$P_{x_{1}}$} and \smash{$P_{x_{2}}$}, \ie the characteristic functions, (ii) taking their logarithms, (iii) using the fact that the logarithms of the characteristic functions are additive, (iv) removing the logarithm, which yields the characteristic function, and so also the moments and the PDF, of the sum \smash{$x_{1} + x_{2}$}. (The logarithm of the characteristic function can be expanded in a power series around $z = 0$, \smash{$r_{x} ( z ) = \sum_{n \geq 1} k_{x , n} z^{n}$}. Its coefficients are called the ``cumulants,'' and are obviously additive, \smash{$k_{x_{1} + x_{2} , n} = k_{x_{1} , n} + k_{x_{2} , n}$}, upon the addition of two independent random variables.)

It is very far from trivial how to extend these steps into the case of random matrices, \ie from the commutative to non--commutative level, and it is the FRV theory of Voiculescu \emph{et al.} and Speicher~\cite{VoiculescuDykemaNica1992,Speicher1994} that develops a precise answer to this question. First of all, FRV puts forth a powerful concept of ``freeness,'' which is a non--commutative analog of independence. We will not delve too deep into explaining its construction, but let us show how it differs from classical independence. Namely, in classical probability, \smash{$x_{1}$} and \smash{$x_{2}$} are independent if their demeaned versions, \smash{$X_{1 , 2} \equiv x_{1 , 2} - \langle x_{1 , 2} \rangle$}, obey \smash{$\langle X_{1} X_{2} \rangle = 0$}. In non--commutative probability, the $m$ non--commutative random variables (random matrices) \smash{$\mathbf{x}_{1} , \ldots , \mathbf{x}_{m}$} are called ``free'' if their demeaned versions \smash{$\mathbf{X}_{j} \equiv \mathbf{x}_{j} - \langle \mathbf{x}_{j} \rangle$} satisfy
\begin{equation}\label{eq:Freeness}
\la p_{1} \left( \mathbf{X}_{j_{1}} \right) \ldots p_{n} \left( \mathbf{X}_{j_{n}} \right) \ra = 0 ,
\end{equation}
for all positive integers $n$, all polynomials \smash{$p_{1} , \ldots , p_{n}$}, and all indices \smash{$j_{1} , \ldots , j_{n} = 1 , \ldots , m$} such that \smash{$j_{1} \neq j_{2} \neq \ldots \neq j_{n}$}. For example, if \smash{$\mathbf{x}_{1}$} and \smash{$\mathbf{x}_{2}$} are free, there will be \smash{$\langle \mathbf{x}_{1}^{2} \mathbf{x}_{2}^{2} \rangle = \langle \mathbf{x}_{1}^{2} \rangle \langle \mathbf{x}_{2}^{2} \rangle$}, \ie just like for independent classical variables, but also \smash{$\langle \mathbf{x}_{1} \mathbf{x}_{2} \mathbf{x}_{1} \mathbf{x}_{2} \rangle = \langle \mathbf{x}_{1}^{2} \rangle \langle \mathbf{x}_{2} \rangle^{2} + \langle \mathbf{x}_{1} \rangle^{2} \langle \mathbf{x}_{2}^{2} \rangle - \langle \mathbf{x}_{1} \rangle^{2} \langle \mathbf{x}_{2} \rangle^{2}$}, much differently than in the commutative situation. In other words, the mixed moments of free non--commutative random variables generally do not factorize into separate moments, as it is the case for independence. Freeness is therefore a much more involved property. To give a practical summary, let us state that random matrices drawn from factorized distributions exhibit (asymptotically, \ie when their sizes tend to infinity) freeness. (Borrowing a picture from physics, we may say that freeness is equivalent to planarity in the limit of a large number of colors in field theory~\cite{CvitanovicLauwersScharbach1982,tHooft1974}.)

Freeness is a relevant idea because the problem of adding two free non--commutative random variables, \smash{$\mathbf{x}_{1} + \mathbf{x}_{2}$}, can be solved in a way analogous to its classical counterpart. Without any proofs (which are not very complicated but lengthy), we will just describe the resulting procedure:
\begin{description}
\item[Step 1:] The moments of the free random matrices, \smash{$\mathbf{x}_{1}$} and \smash{$\mathbf{x}_{2}$}, are conveniently encoded in the Green's functions \smash{$G_{\mathbf{x}_{1}} ( z )$} and \smash{$G_{\mathbf{x}_{2}} ( z )$} (\ref{eq:GreenFunctionDefinition}), (\ref{eq:MomentsExpansionOfTheGreenFunction}).
\item[Step 2:] The Green's functions are inverted functionally to obtain the respective Blue's functions \smash{$B_{\mathbf{x}_{1}} ( z )$} and \smash{$B_{\mathbf{x}_{2}} ( z )$} (\ref{eq:BlueFunctionAndNTransformDefinition}).
\item[Step 3:] The Blue's functions obey the ``non--commutative addition law,''
    \begin{equation}\label{eq:TheNonCommutativeAdditionLaw}
    B_{\mathbf{x}_{1} + \mathbf{x}_{2}} ( z ) = B_{\mathbf{x}_{1}} ( z ) + B_{\mathbf{x}_{2}} ( z ) - \frac{1}{z} , \qquad \textrm{for free \smash{$\mathbf{x}_{1}$}, \smash{$\mathbf{x}_{2}$}.}
    \end{equation}
    (Equivalently, this means that the $R$--transforms are additive, \smash{$R_{\mathbf{x}_{1} + \mathbf{x}_{2}} ( z ) = R_{\mathbf{x}_{1}} ( z ) + R_{\mathbf{x}_{2}} ( z )$}. Trivially, the free cumulants (\ref{eq:FreeCumulantsDefinition}) are additive as well, \smash{$K_{\mathbf{x}_{1} + \mathbf{x}_{2} , n} = K_{\mathbf{x}_{1} , n} + K_{\mathbf{x}_{2} , n}$}. Let us also mention, for the readers familiar with the Feynman diagrammatic techniques, that the additivity of the $R$--transform can be explained in terms of the additivity of the self--energy.)
\item[Step 4:] Invert functionally \smash{$B_{\mathbf{x}_{1} + \mathbf{x}_{2}} ( z )$} to find the Green's function of the sum, \smash{$G_{\mathbf{x}_{1} + \mathbf{x}_{2}} ( z )$}, and subsequently, its mean spectral density \smash{$\rho_{\mathbf{x}_{1} + \mathbf{x}_{2}} ( \lambda )$} through the Sokhotsky formula (\ref{eq:SokhotskyFormula}).
\end{description}

We recognize that it parallels the classical construction: the Green's function is an analog of the characteristic function (\ref{eq:CharacteristicFunctionDefinition}), functional inversion and forming the $R$--transform replaced taking the logarithm (\ref{eq:LogarithmOfTheCharacteristicFunctionDefinition}), and the addition law is a direct generalization of the classical one (\ref{eq:TheClassicalAdditionLaw}). The correspondence between the classical probability calculus and matrix probability calculus (FRV) is thus summarized in the following chart:
\begin{equation}\label{eq:Cartoon}
\begin{array}{ccc} \textrm{PDF} & \qquad \leftrightarrow \qquad & \textrm{spectral density} \\ \downarrow &  & \downarrow \\ \textrm{characteristic function} & \qquad \leftrightarrow \qquad & \textrm{Green's function} \\ \downarrow &  & \downarrow \\ \textrm{logarithm of characteristic function} & \qquad \leftrightarrow \qquad & \textrm{$R$--transform} \\ \downarrow &  & \downarrow \\ \textrm{additivity for independent variables} & \qquad \leftrightarrow \qquad & \textrm{additivity for free variables} \end{array}
\end{equation}

A closely related problem is how to deduce a composition law for the multiplication of free random matrices. The distribution of a product of independent random variables is not widely discussed in textbooks on classical probability theory, since it can be derived from the relation \smash{$\exp x_{1} \exp x_{2} = \exp ( x_{1} + x_{2} )$}, which reduces the multiplication problem to the addition one by a change of variables. However, this is not the case for random matrices, which do not commute: in general, \smash{$\exp \mathbf{x}_{1} \exp \mathbf{x}_{2} \neq \exp ( \mathbf{x}_{1} + \mathbf{x}_{2} )$}. This notwithstanding, there exists~\cite{VoiculescuDykemaNica1992} a transformation (called the ``$S$--transformation'') which allows one to calculate the resolvent of a product of free random matrices \smash{$\mathbf{x}_{1} \mathbf{x}_{2}$} from the resolvents of each separate term, just like there is the $R$--transformation for the sum. Again without proofs, the multiplication algorithm is:
\begin{description}
\item[Step 1:] The moments of the free random matrices, \smash{$\mathbf{x}_{1}$} and \smash{$\mathbf{x}_{2}$}, are conveniently encoded in the moments' generating functions \smash{$M_{\mathbf{x}_{1}} ( z )$} and \smash{$M_{\mathbf{x}_{2}} ( z )$} (\ref{eq:MomentsGeneratingFunctionDefinition}).
\item[Step 2:] The moments' generating functions are inverted functionally to obtain the respective $N$--transforms \smash{$N_{\mathbf{x}_{1}} ( z )$} and \smash{$N_{\mathbf{x}_{2}} ( z )$} (\ref{eq:BlueFunctionAndNTransformDefinition}).
\item[Step 3:] The $N$--transforms obey the ``non--commutative multiplication law,''
    \begin{equation}\label{eq:TheNonCommutativeMultiplicationLaw}
    N_{\mathbf{x}_{1} \mathbf{x}_{2}} ( z ) = \frac{z}{1 + z} N_{\mathbf{x}_{1}} ( z ) N_{\mathbf{x}_{2}} ( z ) , \qquad \textrm{for free \smash{$\mathbf{x}_{1}$}, \smash{$\mathbf{x}_{2}$}.}
    \end{equation}
    (Equivalently, this means that the so--called ``$S$--transforms,'' \smash{$S_{\mathbf{x}} ( z ) \equiv ( 1 + z ) / ( z N_{\mathbf{x}} ( z ) )$}, are multiplicative, \smash{$S_{\mathbf{x}_{1} \mathbf{x}_{2}} ( z ) = S_{\mathbf{x}_{1}} ( z ) S_{\mathbf{x}_{2}} ( z )$}.)
\item[Step 4:] Invert functionally \smash{$N_{\mathbf{x}_{1} \mathbf{x}_{2}} ( z )$} to find the moments' generating function of the product, \smash{$M_{\mathbf{x}_{1} \mathbf{x}_{2}} ( z )$}, and subsequently, its Green's function and mean spectral density.
\end{description}

Let us close with a few comments:
\begin{itemize}
\item There is a one--to--one correspondence between classical and free random variables, which in particular allows one to map probability densities of random variables into the corresponding eigenvalues' densities of large free random matrices~\cite{BercoviciPata1999}.
\item Also, one can define the analog of the concept of stability~\cite{BercoviciVoiculescu1993}, which in the FRV calculus assumes the form of spectral stability.
\item A consequence of the above two observations is that the eigenvalues' distribution of a properly normalized sum of many random matrices for which the second spectral moment is finite tends to a universal limiting distribution known in RMT as Wigner's semicircle law~\cite{Wigner1955}. The Wigner's distribution in the FRV calculus corresponds to the Gaussian distribution in the standard probability calculus.
\item Another consequence is that there exists a counterpart of the L\'{e}vy stable distributions for FRV. Since large random matrices asymptotically represent free random variables, one can expect the existence of large free random matrices in the L\'{e}vy stability class. We will exploit this fact in a forthcoming publication.
\item Recently, it has been proven that FRV exhibits central theorems for extreme values~\cite{BenArousVoiculescu2006}, again in a one--to--one correspondence with the extreme values' distributions known in classical probability from the Fisher--Tippet theorem, \ie the Fr\'{e}chet, Weibull and Gumbel distributions.
\item For completeness, let us also mention that FRV can also generate dynamical stochastic processes~\cite{BianeSpeicher2001,JanikWieczorek2004,GudowskaNowakJanikJurkiewiczNowak2005}, alike Gaussian distributions generate random walks in classical probability. We will not discuss them in this work, restricting our attention to stationary properties of FRV only.
\end{itemize}


\subsection{The Uncorrelated Wishart Ensemble From FRV}
\label{ss:TheUncorrelatedWishartEnsembleFromFRV}


\subsubsection{The Estimator $\mathbf{c}$ for \smash{$\mathbf{C} = \mathbf{1}_{N}$} and \smash{$\mathbf{A} = \mathbf{1}_{T}$}}
\label{sss:TheEstimatorcForCOneAOne}

As a first display of the efficiency of the FRV method, we re--derive the Green's function (equivalently, the moments' generating function; consequently, the density) of the so--called uncorrelated Wishart ensemble~\cite{Wishart1928}, that is, the random matrix $\mathbf{c}$ (\ref{eq:TransformedEstimators}) in which \smash{$\mathbf{C} = \mathbf{1}_{N}$} and \smash{$\mathbf{A} = \mathbf{1}_{T}$} has been set,
\begin{equation}\label{eq:UncorrelatedWishartDefinition}
\mathbf{c} = \frac{1}{T} \widetilde{\mathbf{R}} \widetilde{\mathbf{R}}^{\TT} , \qquad \mathbf{a} = \frac{1}{N} \widetilde{\mathbf{R}}^{\TT} \widetilde{\mathbf{R}} ,
\end{equation}
with the uncorrelated Gaussian probability distribution \smash{$P_{\mathrm{G.}} ( \widetilde{\mathbf{R}} )$} (\ref{eq:UncorrelatedGaussian}). These are the Pearson estimators of the cross--covariance and auto--covariance matrices, respectively, with the trivial underlying covariance structure \smash{$\mathcal{C}_{i a , j b} = \delta_{i j} \delta_{a b}$}. We hope that this short, simple, and entirely algebraic calculation of the result which is relatively well--known in the quantitative finance community (the Mar\v{c}enko--Pastur distribution~\cite{MarcenkoPastur1967}), but found previously only with aid of more involved tools (such as the planar diagrammatic expansion or the replica trick), will convince the reader about the obvious advantages of the FRV calculus.

We will show that the $N$--transform of $\mathbf{c}$, for any value of $r > 0$, reads
\begin{equation}\label{eq:UncorrelatedWishartNTransform}
N_{\mathbf{c}} ( z ) = \frac{( 1 + z ) ( 1 + r z )}{z} ,
\end{equation}
which after functional inversion (solving a quadratic equation; the proper one of the two solutions is chosen so to satisfy (\ref{eq:GreenFunctionAtInfinity}), which implies the minus sign before the principal square root) yields the moments' generating function (which we will not print), and upon using (\ref{eq:MomentsGeneratingFunctionDefinition}), also the Green's function,
\begin{equation}\label{eq:UncorrelatedWishartGreenFunction}
G_{\mathbf{c}} ( z ) = \frac{z + r - 1 - \sqrt{\left( z - \lambda_{+} \right) \left( z - \lambda_{-} \right)}}{2 r z} , \qquad \textrm{where} \qquad \lambda_{\pm} \equiv \left( 1 \pm \sqrt{r} \right)^{2} .
\end{equation}
The Sokhotsky formula (\ref{eq:SokhotskyFormula}) then finally leads to the celebrated Mar\v{c}enko--Pastur spectral density,
\begin{equation}\label{eq:UncorrelatedWishartSpectralDensity}
\rho_{\mathbf{c}} ( \lambda ) = \frac{\sqrt{\left( \lambda_{+} - \lambda \right) \left( \lambda - \lambda_{-} \right)}}{2 \pi r \lambda} , \qquad \textrm{for} \qquad \lambda \in \left[ \lambda_{-} , \lambda_{+} \right] .
\end{equation}


\subsubsection{The Duality}
\label{sss:TheDuality}

Before we proceed to the derivation, it is important to express in a quantitative way the relation (a ``duality'') between $\mathbf{c}$ and $\mathbf{a}$ announced already in par.~\ref{sss:EstimatorsOfCrossCovariances}; this argumentation is valid for arbitrary $\mathbf{C}$ and $\mathbf{A}$. Indeed, the moments satisfy \smash{$M_{\mathbf{c} , n} = r^{n - 1} M_{\mathbf{a} , n}$}, for any $n \geq 1$ and regardless of the value of $r > 0$, due to the cyclic property of the trace (and $M_{\mathbf{c} , 0} = M_{\mathbf{a} , 0} = 1$). In terms of their generating functions, and consequently the Green's functions, this relation reads
\begin{equation}\label{eq:DualityForTheMomentsGeneratingFunctionAndGreenFunction}
M_{\mathbf{a}} ( z ) = r M_{\mathbf{c}} ( r z ) , \qquad \textrm{or equivalently} \qquad G_{\mathbf{a}} ( z ) = r^{2} G_{\mathbf{c}} ( r z ) + \frac{1 - r}{z} .
\end{equation}
These formulae can obviously be inverted to yield $\mathbf{c}$ in terms of $\mathbf{a}$: it amounts to simultaneously exchanging $\mathbf{c} \leftrightarrow \mathbf{a}$, $\mathbf{C} \leftrightarrow \mathbf{A}$ and $r \leftrightarrow 1 / r$. Also, they remain intact even when the measure is not Gaussian, and even when the moments do not exist; in this case, a proof features a simple algebraic manipulation using the definition of the Green's function (\ref{eq:GreenFunctionDefinition}). As mentioned before, (\ref{eq:DualityForTheMomentsGeneratingFunctionAndGreenFunction}) means that the information carried by $\mathbf{c}$ and $\mathbf{a}$ is equivalent, and we may safely forget about one of them, say $\mathbf{a}$. Also, we will use (\ref{eq:DualityForTheMomentsGeneratingFunctionAndGreenFunction}) in the following.


\subsubsection{An FRV Derivation of (\ref{eq:UncorrelatedWishartNTransform})}
\label{sss:TheUncorrelatedWishartEnsembleFromFRVAnFRVDerivation}

We will now present a purely algebraic computation~\cite{JanikNowakPappWambachZahed1997,Voiculescu1991} of the $N$--transforms of both the uncorrelated Wishart matrices $\mathbf{c}$ and $\mathbf{a}$ (\ref{eq:UncorrelatedWishartDefinition}) based on the multiplication property of the $N$--transform for free random matrices (\ref{eq:TheNonCommutativeMultiplicationLaw}).

It is convenient to start from assuming $N \leq T$ (\ie $r \leq 1$) and considering the random $T \times T$ matrix \smash{$( 1 / T ) \widetilde{\mathbf{R}}^{\TT} \widetilde{\mathbf{R}}$}. The following trick is exploited in order to work with square matrices instead of rectangular: We introduce a square $T \times T$ random matrix $\mathbf{X}$ with real uncorrelated Gaussian entries, \smash{$P_{\mathrm{G.}} ( \mathbf{X} ) \propto \exp ( - ( 1 / 2 ) \sum_{a b} X_{a b}^{2} ) = \exp ( - ( 1 / 2 ) \Tr \mathbf{X}^{\TT} \mathbf{X} )$}. Next, we use the projector
\begin{equation}\label{eq:ProjectorDefinition}
\mathbf{P} \equiv \diag \left( \mathbf{1}_{N} , \mathbf{0}_{T - N} \right) ,
\end{equation}
to cut from $\mathbf{X}$ an $N \times T$ rectangle, \smash{$\widetilde{\mathbf{R}}_{0} \equiv \mathbf{P} \mathbf{X}$}. More precisely, this is a square $T \times T$ matrix whose all the entries are zero but the ``upper'' $N \times T$ rectangle. This rectangle may be called \smash{$\widetilde{\mathbf{R}}$}, since all its entries are uncorrelated Gaussian random variables. Hence,
\begin{equation}\label{eq:TheUncorrelatedWishartEnsembleFromFRVAnFRVDerivation1}
\frac{1}{T} \widetilde{\mathbf{R}}^{\TT} \widetilde{\mathbf{R}} = \frac{1}{T} \widetilde{\mathbf{R}}_{0}^{\TT} \widetilde{\mathbf{R}}_{0} = \frac{1}{T} \mathbf{X}^{\TT} \mathbf{P} \mathbf{X} .
\end{equation}

Furthermore, thanks to the cyclic property of the trace, all the moments of this matrix are equal to the moments of \smash{$( 1 / T ) \mathbf{P} \mathbf{X} \mathbf{X}^{\TT}$}, so also their $N$--transforms coincide. Now, this is a product of two free matrices, $\mathbf{P}$ and \smash{$( 1 / T ) \mathbf{X} \mathbf{X}^{\TT}$}, therefore, the multiplication law (\ref{eq:TheNonCommutativeMultiplicationLaw}) allows to write
\begin{equation}\label{eq:TheUncorrelatedWishartEnsembleFromFRVAnFRVDerivation2}
N_{\frac{1}{T} \mathbf{X}^{\TT} \mathbf{P} \mathbf{X}} ( z ) = N_{\frac{1}{T} \mathbf{P} \mathbf{X} \mathbf{X}^{\TT}} ( z ) = \frac{z}{1 + z} N_{\mathbf{P}} ( z ) N_{\frac{1}{T} \mathbf{X} \mathbf{X}^{\TT}} ( z ) .
\end{equation}

The $N$--transform of the projector is easily computed,
\begin{equation}\label{eq:ProjectorNTransform}
N_{\mathbf{P}} ( z ) = 1 + \frac{r}{z} ,
\end{equation}
because all its moments \smash{$M_{\mathbf{P} , n} = ( 1 / T ) \Tr \mathbf{P}^{n} = ( 1 / T ) \Tr \mathbf{P} = r$}, $n \geq 1$, hence \smash{$M_{\mathbf{P}} ( z ) = r / ( z - 1 )$}, whose functional inversion is the above.

It remains therefore to find the $N$--transform of \smash{$( 1 / T ) \mathbf{X} \mathbf{X}^{\TT}$}. We recognize that this is an uncorrelated Wishart random matrix with $r = 1$; in other words, the projector trick and the FRV multiplication law reduced the problem with an arbitrary $r$ to solving the $r = 1$ case. Now, this simplified problem is handled by noticing that the spectral properties of the $r = 1$ Wishart ensemble are equivalent to that of the squared Gaussian Orthogonal Ensemble (GOE). The argumentation is more clear in the case of complex entries in $\mathbf{X}$; and at the leading order in the large--$T$ limit there is no difference between the real and complex versions. Namely, $\mathbf{X}$ can be decomposed as a sum of its Hermitian and anti--Hermitian parts, \smash{$\mathbf{X} = \mathbf{H}_{1} + \ii \mathbf{H}_{2}$}, which implies \smash{$\Tr \mathbf{X} \mathbf{X}^{\dagger} = \Tr ( \mathbf{H}_{1}^{2} + \mathbf{H}_{2}^{2} )$}. This means that the Gaussian measure for $\mathbf{X}$ factorizes, \ie \smash{$\mathbf{H}_{1}$} and \smash{$\mathbf{H}_{2}$} are two independent Hermitian random matrices (GUEs). More generally, \smash{$\Tr ( \mathbf{X} \mathbf{X}^{\dagger} )^{n} = \Tr ( \mathbf{H}_{1}^{2} + \mathbf{H}_{2}^{2} )^{n}$}, for any integer $n \geq 1$, hence the random matrix \smash{$\mathbf{X} \mathbf{X}^{\dagger}$} is equivalent to a sum of two squared GUEs. Returning to real matrices, and taking into account the corresponding rescaling of the variance, we arrive at the conclusion that
\begin{equation}\label{eq:TheUncorrelatedWishartEnsembleFromFRVAnFRVDerivation3}
N_{\frac{1}{T} \mathbf{X} \mathbf{X}^{\TT}} ( z ) = N_{\mathbf{GOE}^{2}} ( z ) .
\end{equation}
The spectral properties of the square of a matrix are related to those of the matrix by a simple algebraic manipulation, \smash{$1 / ( z^{2} \mathbf{1}_{T} - \mathbf{H}^{2} ) = ( 1 / ( z \mathbf{1}_{T} - \mathbf{H} ) + 1 / ( z \mathbf{1}_{T} + \mathbf{H} ) ) / 2 z$}, which implies, in the relevant situation when all the odd moments vanish,
\begin{equation}\label{eq:TheUncorrelatedWishartEnsembleFromFRVAnFRVDerivation4}
M_{\mathbf{H}^{2}} \left( z^{2} \right) = M_{\mathbf{H}} ( z ) .
\end{equation}
The moments' generating function of the GOE is well--known and given by the Wigner's formula~\cite{Wigner1955},
\begin{equation}\label{eq:TheUncorrelatedWishartEnsembleFromFRVAnFRVDerivation5}
M_{\mathbf{GOE}} ( z ) = \frac{z}{2} \left( z - \sqrt{z^{2} - 4} \right) - 1 .
\end{equation}
These ingredients (\ref{eq:TheUncorrelatedWishartEnsembleFromFRVAnFRVDerivation3}), (\ref{eq:TheUncorrelatedWishartEnsembleFromFRVAnFRVDerivation4}), (\ref{eq:TheUncorrelatedWishartEnsembleFromFRVAnFRVDerivation5}) assembled together lead to the $N$--transform of the $r = 1$ uncorrelated Wishart,
\begin{equation}\label{eq:SquareUncorrelatedWishartNTransform}
N_{\frac{1}{T} \mathbf{X} \mathbf{X}^{\TT}} ( z ) = \frac{( 1 + z )^{2}}{z} .
\end{equation}

Plugging (\ref{eq:ProjectorNTransform}) and (\ref{eq:SquareUncorrelatedWishartNTransform}) into (\ref{eq:TheUncorrelatedWishartEnsembleFromFRVAnFRVDerivation2}), and using (\ref{eq:TheUncorrelatedWishartEnsembleFromFRVAnFRVDerivation1}), finally yields
\begin{equation}\label{eq:TheUncorrelatedWishartEnsembleResult1}
N_{\frac{1}{T} \widetilde{\mathbf{R}}^{\TT} \widetilde{\mathbf{R}}} ( z ) = \frac{( 1 + z ) ( r + z )}{z} ,
\end{equation}
which we recall has been derived for $r \leq 1$. The scaling relation \smash{$N_{g \mathbf{H}} ( z ) = g N_{\mathbf{H}} ( z )$}, true for any random matrix $\mathbf{H}$ and non--zero complex constant $g$, implies further that
\begin{equation}\label{eq:TheUncorrelatedWishartEnsembleResult2}
N_{\frac{1}{N} \widetilde{\mathbf{R}}^{\TT} \widetilde{\mathbf{R}}} ( z ) = \frac{( 1 + z ) ( r + z )}{r z} .
\end{equation}
Moreover, the cyclic property of the trace applied in these formulae provides us with
\begin{equation}\label{eq:TheUncorrelatedWishartEnsembleResult3}
N_{\frac{1}{T} \widetilde{\mathbf{R}} \widetilde{\mathbf{R}}^{\TT}} ( z ) = \frac{( 1 + z ) ( 1 + r z )}{z} ,
\end{equation}
and
\begin{equation}\label{eq:TheUncorrelatedWishartEnsembleResult4}
N_{\frac{1}{N} \widetilde{\mathbf{R}} \widetilde{\mathbf{R}}^{\TT}} ( z ) = \frac{( 1 + z ) ( 1 + r z )}{r z} .
\end{equation}
Although we originally assumed that $r \leq 1$, we observe that (\ref{eq:TheUncorrelatedWishartEnsembleResult1}) and (\ref{eq:TheUncorrelatedWishartEnsembleResult4}) transform into each other as we exchange $r \leftrightarrow 1 / r$ and \smash{$\widetilde{\mathbf{R}} \leftrightarrow \widetilde{\mathbf{R}}^{\TT}$}, as do (\ref{eq:TheUncorrelatedWishartEnsembleResult2}) and (\ref{eq:TheUncorrelatedWishartEnsembleResult3}); this is precisely the duality (\ref{eq:DualityForTheMomentsGeneratingFunctionAndGreenFunction}). It means that all these results hold true for any $r > 0$. This completes the derivation, since (\ref{eq:TheUncorrelatedWishartEnsembleResult3}) is the desired $N$--transform of $\mathbf{c}$ (\ref{eq:UncorrelatedWishartNTransform}).


\subsection{The Doubly Correlated Wishart Ensemble From FRV}
\label{ss:TheDoublyCorrelatedWishartEnsembleFromFRV}


\subsubsection{The Estimator $\mathbf{c}$ for Arbitrary \smash{$\mathbf{C}$} and \smash{$\mathbf{A}$} (the Main Result)}
\label{sss:TheEstimatorcForArbitraryCAndATheMainResult}

In this subsection, which constitutes the central piece of our work, we will consider the doubly correlated Wishart random matrix \smash{$\mathbf{c} = ( 1 / T ) \sqrt{\mathbf{C}} \widetilde{\mathbf{R}} \mathbf{A} \widetilde{\mathbf{R}}^{\TT} \sqrt{\mathbf{C}}$} (\ref{eq:TransformedEstimators}), as well as its time--lagged version \smash{$\mathbf{c}^{\mathrm{sym.} ( d )} = ( 1 / T ) \sqrt{\mathbf{C}} \widetilde{\mathbf{R}} \sqrt{\mathbf{A}} \mathbf{D}^{\mathrm{sym.} ( d )} \sqrt{\mathbf{A}} \widetilde{\mathbf{R}}^{\TT} \sqrt{\mathbf{C}}$} (\ref{eq:TransformedSymmetrizedLaggedCovarianceMatrixEstimator}), and show how a back--of--an--envelope calculation, founded upon the FRV multiplication law (\ref{eq:TheNonCommutativeMultiplicationLaw}), leads to expressions for the $N$--transforms of these estimators; through functional inversions, these expressions will yield equations for the moments' generating functions of $\mathbf{c}$ and \smash{$\mathbf{c}^{\mathrm{sym.} ( d )}$}, which in turn carry the full information about the spectral properties of these estimators.

The $N$--transform of the estimator $\mathbf{c}$ in the case of arbitrary underlying covariance matrices will be found in par.~\ref{sss:TheDoublyCorrelatedWishartEnsembleFromFRVAnFRVDerivation} to be
\begin{equation}\label{eq:DoublyCorrelatedWishartNTransform}
N_{\mathbf{c}} ( z ) = r z N_{\mathbf{A}} ( r z ) N_{\mathbf{C}} ( z ) .
\end{equation}
In other words, this is an equation for the moments' generating function \smash{$M \equiv M_{\mathbf{c}} ( z )$},
\begin{equation}\label{eq:DoublyCorrelatedWishartEquationForMomentsGeneratingFunction}
z = r M N_{\mathbf{A}} ( r M ) N_{\mathbf{C}} ( M ) .
\end{equation}

A few comments are in place:
\begin{itemize}
\item When $\mathbf{C}$ is arbitrary, but there are no auto--covariances, \smash{$\mathbf{A} = \mathbf{1}_{T}$}, we have \smash{$N_{\mathbf{A}} ( z ) = 1 + 1 / z$}, hence equation (\ref{eq:DoublyCorrelatedWishartEquationForMomentsGeneratingFunction}) becomes
    \begin{equation}\label{eq:DoublyCorrelatedWishartEquationForMomentsGeneratingFunctionCArbitraryAOne}
    M = M_{\mathbf{C}} \left( \frac{z}{1 + r M} \right) .
    \end{equation}
\item A similar simplification occurs when $\mathbf{A}$ is arbitrary, but there are no cross--covariances, \smash{$\mathbf{C} = \mathbf{1}_{N}$}, in which case
    \begin{equation}\label{eq:DoublyCorrelatedWishartEquationForMomentsGeneratingFunctionCOneAArbitrary}
    r M = M_{\mathbf{A}} \left( \frac{z}{r ( 1 + M )} \right) .
    \end{equation}
\end{itemize}


\subsubsection{The Estimator \smash{$\mathbf{c}^{\mathrm{sym.} ( d )}$} for Arbitrary $\mathbf{C}$ and $\mathbf{A}$}
\label{sss:TheEstimatorcForArbitraryCAndALaggedVersion}

A one--line computation presented in par.~\ref{sss:TheDoublyCorrelatedWishartEnsembleFromFRVAnFRVDerivationLaggedVersion} leads from (\ref{eq:DoublyCorrelatedWishartNTransform}) to a formula for the $N$--transform of the time--lagged estimator \smash{$\mathbf{c}^{\mathrm{sym.} ( d )}$}, since the latter is a version of the former with a modified auto--covariance matrix $\mathbf{A}$,
\begin{equation}\label{eq:DoublyCorrelatedWishartNTransformLaggedVersion}
N_{\mathbf{c}^{\mathrm{sym.} ( d )}} ( z ) = \frac{r^{2} z^{2}}{1 + r z} N_{\mathbf{D}^{\mathrm{sym.} ( d )}} ( r z ) N_{\mathbf{A}} ( r z ) N_{\mathbf{C}} ( z ) .
\end{equation}
Equivalently, this is an equation obeyed by the moments' generating function \smash{$M \equiv M_{\mathbf{c}^{\mathrm{sym.} ( d )}} ( z )$},
\begin{equation}\label{eq:DoublyCorrelatedWishartEquationForMomentsGeneratingFunctionLaggedVersion}
z = \frac{r^{2} M^{2}}{1 + r M} N_{\mathbf{D}^{\mathrm{sym.} ( d )}} ( r M ) N_{\mathbf{A}} ( r M ) N_{\mathbf{C}} ( M ) ,
\end{equation}

In par.~\ref{sss:TheDoublyCorrelatedWishartEnsembleFromFRVAnFRVDerivationLaggedVersion} we derive an explicit expression for the Green's function of the symmetrized delay matrix \smash{$\mathbf{D}^{\mathrm{sym.} ( d )}$} (\ref{eq:SymmetrizedLaggedCovarianceMatrixEstimator}), which allows to find its $N$--transform. Recalling that $t \equiv T / d$ is an integer $\geq 2$, there is
\begin{equation}\label{eq:GreenFunctionForTheSymmetrizedDelayMatrix}
G_{\mathbf{D}^{\mathrm{sym.} ( d = T / t )}} ( z ) = \frac{1}{t} \sum_{\dot{a} = 1}^{t} \frac{1}{z - \cos \frac{\pi \dot{a}}{t + 1}} = \left\{ \begin{array}{ll} \frac{2 z}{t} \sum_{l = 1}^{t / 2} \frac{1}{z^{2} - \cos^{2} \frac{\pi l}{t + 1}} & \qquad \textrm{for $t$ even,} \\ \frac{1}{t z} + \frac{2 z}{t} \sum_{l = 1}^{( t - 1 ) / 2} \frac{1}{z^{2} - \cos^{2} \frac{\pi l}{t + 1}} & \qquad \textrm{for $t$ odd.} \end{array} \right.
\end{equation}
Notice that this result (\ref{eq:GreenFunctionForTheSymmetrizedDelayMatrix}) does not depend on $T$ or $d$ separately, but only on their ratio $t$. In particular, it remains true in the limit
\begin{equation}\label{eq:ThermodynamicalLimitLaggedVersionOne}
T \to \infty , \qquad d \to \infty , \qquad \textrm{such that} \qquad t = \textrm{fixed} ,
\end{equation}
in which we have a very long time series divided into a fixed number of very long lags. Such a situation may be financially relevant: Indeed, a natural choice for the lag $d$ would be the scale $\tau$ of the true temporal correlations existing in the system (in the units of $\delta t$). And for example, if one assumes that the proper description of heteroscedasticity is through the I--GARCH$( 1 )$ process with the parameter $\alpha$ (see par.~\ref{sss:Heteroscedasticity}), there appears a characteristic time $\tau = - 1 / \log \alpha$. A financially justified limit (\ref{eq:EWMALimit}), which we discuss later, can then be taken in which $\tau \sim T$; hence, (\ref{eq:ThermodynamicalLimitLaggedVersionOne}) seems to be able to probe a relevant regime.

Another interesting limit would be of a very long time series with a finite time lag,
\begin{equation}\label{eq:ThermodynamicalLimitLaggedVersionTwo}
T \to \infty , \qquad t \to \infty , \qquad \textrm{such that} \qquad d = \textrm{fixed} ,
\end{equation}
in which case the sum in (\ref{eq:GreenFunctionForTheSymmetrizedDelayMatrix}) can be approximated by an integral,
\begin{equation}\label{eq:GreenFunctionForTheSymmetrizedDelayMatrixNearestNeighbor}
G_{\mathbf{D}^{\mathrm{sym.} ( d = \textrm{fixed} )}} ( z ) = \int_{0}^{1} \dd x \frac{1}{z - \cos ( \pi x )} = \frac{1}{\sqrt{z^{2} - 1}} , \qquad \textrm{hence,} \qquad N_{\mathbf{D}^{\mathrm{sym.} ( d = \textrm{fixed} )}} ( z ) = \frac{1 + z}{\sqrt{z ( 2 + z )}} .
\end{equation}
(This can be checked to be equivalent to the infinite symmetrized delay matrix with $d = 1$, \ie with the ``nearest--neighbor'' delay. A finite $d$ compared to an infinite $T$ is just like $d = 1$. We remark that (\ref{eq:GreenFunctionForTheSymmetrizedDelayMatrixNearestNeighbor}) may as well be obtained through the method sketched in par.~\ref{sss:IntroductionInfiniteTranslationallyInvariantMatrices}.) Equation (\ref{eq:DoublyCorrelatedWishartEquationForMomentsGeneratingFunctionLaggedVersion}) acquires the form
\begin{equation}\label{eq:DoublyCorrelatedWishartEquationForMomentsGeneratingFunctionLaggedVersionNearestNeighbor}
z = r M \sqrt{\frac{r M}{2 + r M}} N_{\mathbf{A}} ( r M ) N_{\mathbf{C}} ( M ) .
\end{equation}
Formally, it is equivalent to the corresponding one for the usual estimator $\mathbf{c}$ (\ref{eq:DoublyCorrelatedWishartEquationForMomentsGeneratingFunction}) with the substitution \smash{$z \to z \sqrt{1 + 2 / ( r M )}$}. Let us however print some of its special cases:
\begin{itemize}
\item When there are no underlying covariances, \smash{$\mathbf{C} = \mathbf{1}_{N}$} and \smash{$\mathbf{A} = \mathbf{1}_{T}$}, (\ref{eq:DoublyCorrelatedWishartEquationForMomentsGeneratingFunctionLaggedVersionNearestNeighbor}) becomes a fourth--order polynomial (Ferrari) equation for $M$,
    \begin{equation}\label{eq:DoublyCorrelatedWishartEquationForMomentsGeneratingFunctionLaggedVersionNearestNeighborCOneAOne}
    r^{2} M^{4} + 2 r ( 1 + r ) M^{3} + \left( 1 + 4 r + r^{2} - z^{2} \right) M^{2} + 2 \left( 1 + r - \frac{z^{2}}{r} \right) M + 1 = 0 .
    \end{equation}
    It coincides with the result presented without proof in~\cite{MayyaAmritkar2006}.
\item For $\mathbf{C}$ arbitrary and \smash{$\mathbf{A} = \mathbf{1}_{T}$},
    \begin{equation}\label{eq:DoublyCorrelatedWishartEquationForMomentsGeneratingFunctionLaggedVersionNearestNeighborCArbitraryAOne}
    M = M_{\mathbf{C}} \left( \frac{z}{1 + r M} \sqrt{1 + \frac{2}{r M}} \right) .
    \end{equation}
\item For $\mathbf{A}$ is arbitrary and \smash{$\mathbf{C} = \mathbf{1}_{N}$},
    \begin{equation}\label{eq:DoublyCorrelatedWishartEquationForMomentsGeneratingFunctionLaggedVersionNearestNeighborCOneAArbitrary}
    r M = M_{\mathbf{A}} \left( \frac{z}{r ( 1 + M )} \sqrt{1 + \frac{2}{r M}} \right) .
    \end{equation}
\end{itemize}


\subsubsection{An FRV Derivation of (\ref{eq:DoublyCorrelatedWishartNTransform})}
\label{sss:TheDoublyCorrelatedWishartEnsembleFromFRVAnFRVDerivation}

The idea behind the following proof is to reduce the problem in the case of arbitrary underlying covariance matrices \smash{$\mathbf{C}$} and \smash{$\mathbf{A}$} to the uncorrelated version (solved in par.~\ref{sss:TheUncorrelatedWishartEnsembleFromFRVAnFRVDerivation}) by successive use of the cyclic property of the trace and the FRV multiplication formula for the $N$--transforms (\ref{eq:TheNonCommutativeMultiplicationLaw}). Indeed, the cyclic property allows to write
\begin{equation}\label{eq:TheDoublyCorrelatedWishartEnsembleFromFRVAnFRVDerivationDerivation1}
N_{\mathbf{c}} ( z ) = N_{\frac{1}{T} \widetilde{\mathbf{R}} \mathbf{A} \widetilde{\mathbf{R}}^{\TT} \mathbf{C}} ( z ) = \ldots .
\end{equation}
Being a product of two free random matrices, the multiplication law gives further
\begin{equation}\label{eq:TheDoublyCorrelatedWishartEnsembleFromFRVAnFRVDerivationDerivation2}
\ldots = \frac{z}{1 + z} N_{\frac{1}{T} \widetilde{\mathbf{R}} \mathbf{A} \widetilde{\mathbf{R}}^{\TT}} ( z ) N_{\mathbf{C}} ( z ) = \ldots .
\end{equation}
Again, the cyclic property applied to the first of these matrices implies
\begin{equation}\label{eq:TheDoublyCorrelatedWishartEnsembleFromFRVAnFRVDerivationDerivation3}
\ldots = \frac{z}{1 + z} N_{\frac{1}{T} \widetilde{\mathbf{R}}^{\TT} \widetilde{\mathbf{R}} \mathbf{A}} ( r z ) N_{\mathbf{C}} ( z ) = \ldots ,
\end{equation}
where the argument $r z$ appeared because the cyclic shift changed an $N \times N$ matrix into a $T \times T$ one, which accordingly rescaled the moments. The first $N$--transform here is of a product of two free random matrices, hence further
\begin{equation}\label{eq:TheDoublyCorrelatedWishartEnsembleFromFRVAnFRVDerivationDerivation4}
\ldots = \frac{z}{1 + z} \frac{r z}{1 + r z} N_{\frac{1}{T} \widetilde{\mathbf{R}}^{\TT} \widetilde{\mathbf{R}}} ( r z ) N_{\mathbf{A}} ( r z ) N_{\mathbf{C}} ( z ) = \ldots .
\end{equation}
In this way, exploiting twice the cyclic property of the trace and twice the FRV multiplication law, the problem has been boiled down to the uncorrelated case, solved in (\ref{eq:TheUncorrelatedWishartEnsembleResult1}), which finally produces the announced result (\ref{eq:DoublyCorrelatedWishartNTransform}),
\begin{equation}\label{eq:TheDoublyCorrelatedWishartEnsembleFromFRVAnFRVDerivationDerivation5}
\ldots = r z N_{\mathbf{A}} ( r z ) N_{\mathbf{C}} ( z ) ,
\end{equation}
equivalent to equation (\ref{eq:DoublyCorrelatedWishartEquationForMomentsGeneratingFunction}) for \smash{$M_{\mathbf{c}} ( z )$}.

Let us make a few comments:
\begin{itemize}
\item The method is remarkably simpler than other known approaches (planar Feynman diagrams, the replica trick). Equation (\ref{eq:DoublyCorrelatedWishartEquationForMomentsGeneratingFunction}) has been found through diagrammatics in~\cite{BurdaJurkiewiczWaclaw2005-1,BurdaJurkiewiczWaclaw2005-2}, and even earlier, in the case of \smash{$\mathbf{A} = \mathbf{1}_{T}$}, in~\cite{BurdaGorlichJaroszJurkiewicz2004,BurdaJurkiewicz2004}.
\item It does not rely on the existence of the moments.
\item It is not specified to Gaussian randomness. In particular, it may be extended to the more general instance of the L\'{e}vy randomness.
\item It can be generalized to longer strings of free random matrices.
\item It is valid (as is the entire FRV calculus) only in the thermodynamical limit (\ref{eq:ThermodynamicalLimit}) of $N$, $T$ large with $r = N / T$ fixed. For finite values of $N$, $T$, there will in general be finite--size corrections \smash{$\mathrm{O} ( 1 / N^{p} )$}, where $p$ depends on the type of randomness.
\end{itemize}


\subsubsection{An FRV Derivation of (\ref{eq:DoublyCorrelatedWishartNTransformLaggedVersion}) and (\ref{eq:GreenFunctionForTheSymmetrizedDelayMatrix})}
\label{sss:TheDoublyCorrelatedWishartEnsembleFromFRVAnFRVDerivationLaggedVersion}

As stated in par.~\ref{sss:EstimatorsOfTimeDelayedCrossCovariances}, the estimator of the symmetrized time--delayed cross--covariance matrix \smash{$\mathbf{c}^{\mathrm{sym.} ( d )}$} (\ref{eq:TransformedSymmetrizedLaggedCovarianceMatrixEstimator}) has the same form as the usual estimator $\mathbf{c}$ (\ref{eq:TransformedEstimators}), only with a modified true auto--covariance matrix, \smash{$\mathbf{A} \to \sqrt{\mathbf{A}} \mathbf{D}^{\mathrm{sym.} ( d )} \sqrt{\mathbf{A}}$}. Therefore, the formula (\ref{eq:DoublyCorrelatedWishartNTransformLaggedVersion}) for the $N$--transform of \smash{$\mathbf{c}^{\mathrm{sym.} ( d )}$} is proven by using the result (\ref{eq:DoublyCorrelatedWishartNTransform}) for $\mathbf{c}$ with this modification included. Now, the $N$--transform for the modified underlying auto--covariance matrix is obtained through the cyclic property of the trace and the FRV multiplication law,
\begin{equation}\label{eq:TheDoublyCorrelatedWishartEnsembleFromFRVAnFRVDerivationLaggedVersionDerivation1}
N_{\sqrt{\mathbf{A}} \mathbf{D}^{\mathrm{sym.} ( d )} \sqrt{\mathbf{A}}} ( z ) = N_{\mathbf{D}^{\mathrm{sym.} ( d )} \mathbf{A}} ( z ) = \frac{z}{1 + z} N_{\mathbf{D}^{\mathrm{sym.} ( d )}} ( z ) N_{\mathbf{A}} ( z ) ,
\end{equation}
which readily justifies (\ref{eq:DoublyCorrelatedWishartNTransformLaggedVersion}).

The obstacle we are facing at this point is to evaluate the $N$--transform of the symmetrized delay matrix \smash{$D^{\mathrm{sym.} ( d )}_{a b} = ( 1 / 2 ) ( \delta_{a + d , b} + \delta_{a - d , b} )$} (\ref{eq:SymmetrizedLaggedCovarianceMatrixEstimator}). This is a symmetric $T \times T$ matrix, and we recall that the lag $d$ is an integer such that $t \equiv T / d$ is an integer $\geq 2$; as for now, these numbers are finite. For this purpose, the delay matrix should be diagonalized.

First, we remark that \smash{$\mathbf{D}^{\mathrm{sym.} ( d )}$} can be regarded as a $t \times t$ block matrix, with blocks of size $d \times d$, each proportional to the unit matrix \smash{$\mathbf{1}_{d}$}, and the block matrix having the structure of the so--called ``nearest--neighbor delay matrix,'' which is \smash{$\mathbf{D}^{\mathrm{sym.} ( d = 1 )}$} but of size $t \times t$, \smash{$D^{\mathrm{n.n.} ( t )}_{\dot{a} \dot{b}} \equiv ( 1 / 2 ) ( \delta_{\dot{a} + 1 , \dot{b}} + \delta_{\dot{a} - 1 , \dot{b}} )$}, \smash{$\dot{a} , \dot{b} = 1 , \ldots , t$}. Concisely,
\begin{equation}\label{eq:TheDoublyCorrelatedWishartEnsembleFromFRVAnFRVDerivationLaggedVersionDerivation2}
\mathbf{D}^{\mathrm{sym.} ( d )} = \mathbf{D}^{\mathrm{n.n.} ( t )} \otimes \mathbf{1}_{d} .
\end{equation}
We infer from (\ref{eq:TheDoublyCorrelatedWishartEnsembleFromFRVAnFRVDerivationLaggedVersionDerivation2}) that the eigenvalues of \smash{$\mathbf{D}^{\mathrm{sym.} ( d )}$} are just the eigenvalues of \smash{$\mathbf{D}^{\mathrm{n.n.} ( t )}$}, denote them by $\lambda^{\mathrm{n.n.} ( t )}_{\dot{a}}$, each one taken $d$ times. Consequently,
\begin{equation}\label{eq:TheDoublyCorrelatedWishartEnsembleFromFRVAnFRVDerivationLaggedVersionDerivation3}
G_{\mathbf{D}^{\mathrm{sym.} ( d )}} ( z ) = \frac{1}{t d} \sum_{\dot{a} = 1}^{t} \frac{d}{z - \lambda^{\mathrm{n.n.} ( t )}_{\dot{a}}} = \frac{1}{t} \sum_{\dot{a} = 1}^{t} \frac{1}{z - \lambda^{\mathrm{n.n.} ( t )}_{\dot{a}}} = G_{\mathbf{D}^{\mathrm{n.n.} ( t )}} ( z ) .
\end{equation}
\ie the two Green's functions are equal. The task is thus reduced to diagonalizing the $t \times t$ nearest--neighbor delay matrix.

This can be done analytically. For simplicity, consider the nearest--neighbor delay matrix without the prefactor $1 / 2$, \smash{$2 \mathbf{D}^{\mathrm{n.n.} ( t )}$}. Its characteristic determinant, \smash{$\mathcal{D}^{( t )} ( \gamma ) \equiv \Det ( 2 \mathbf{D}^{\mathrm{n.n.} ( t )} - \gamma \mathbf{1}_{d} )$}, is straightforwardly computed inductively w.r.t. $t$ by expanding w.r.t. the first row, \smash{$\mathcal{D}^{( t )} ( \gamma ) = - \gamma \mathcal{D}^{( t - 1 )} ( \gamma ) - \mathcal{D}^{( t - 2 )} ( \gamma )$}, for $t \geq 2$, where we assume \smash{$\mathcal{D}^{( 0 )} ( \gamma ) \equiv 1$}. This recurrence relation (which generates the Fibonacci series) can be solved for example by the generating function technique~\cite{GrahamKnuthPatashnik1994}, and gives
\begin{equation}\label{eq:TheDoublyCorrelatedWishartEnsembleFromFRVAnFRVDerivationLaggedVersionDerivation4}
\mathcal{D}^{( t )} ( \gamma ) = \frac{1}{s_{2} - s_{1}} \left( \frac{1}{s_{1}^{t + 1}} - \frac{1}{s_{2}^{t + 1}} \right) ,
\end{equation}
where \smash{$s_{1}$}, \smash{$s_{2}$} are the two roots of the quadratic equation \smash{$1 + \gamma s + s^{2} = 0$}, and we must constrain \smash{$s_{1} \neq s_{2}$} (\ie $| \gamma | \neq 2$), since it can be verified that otherwise there are no solutions to the characteristic equation. The characteristic equation \smash{$\mathcal{D}^{( t )} ( \gamma ) = 0$} is therefore equivalent to
\begin{equation}\label{eq:TheDoublyCorrelatedWishartEnsembleFromFRVAnFRVDerivationLaggedVersionDerivation5}
s_{1}^{t + 1} = s_{2}^{t + 1} .
\end{equation}
If \smash{$s_{1}$}, \smash{$s_{2}$} were real (\ie $| \gamma | > 2$), this would imply \smash{$s_{1} = s_{2}$}, which is impossible. Hence, there must be $| \gamma | < 2$, and \smash{$s_{1}$}, \smash{$s_{2}$} complex and mutually conjugate,
\begin{equation}\label{eq:TheDoublyCorrelatedWishartEnsembleFromFRVAnFRVDerivationLaggedVersionDerivation6}
s_{1 , 2} = - \frac{\gamma}{2} \pm \ii \frac{\sqrt{4 - \gamma^{2}}}{2} = \ee^{\pm \ii \phi} , \qquad \textrm{where} \qquad \tan \phi = - \frac{\sqrt{4 - \gamma^{2}}}{\gamma} , \qquad \phi \in [ - \pi , \pi ) .
\end{equation}
Then (\ref{eq:TheDoublyCorrelatedWishartEnsembleFromFRVAnFRVDerivationLaggedVersionDerivation5}) means that \smash{$s_{1} / s_{2}$} is a $( t + 1 )$--th root of unity; there are $( t + 1 )$ such roots, but we must exclude the one equal to $1$, so there remain $t$ roots, and the characteristic equation becomes
\begin{equation}\label{eq:TheDoublyCorrelatedWishartEnsembleFromFRVAnFRVDerivationLaggedVersionDerivation7}
\frac{s_{1}}{s_{2}} = \exp \frac{2 \pi \ii \dot{a}}{t + 1} , \qquad \textrm{for} \qquad \dot{a} = 1 , \ldots , t .
\end{equation}
Comparing (\ref{eq:TheDoublyCorrelatedWishartEnsembleFromFRVAnFRVDerivationLaggedVersionDerivation6}) and (\ref{eq:TheDoublyCorrelatedWishartEnsembleFromFRVAnFRVDerivationLaggedVersionDerivation7}) finally provides the eigenvalues of the $t \times t$ nearest--neighbor delay matrix,
\begin{equation}\label{eq:TheDoublyCorrelatedWishartEnsembleFromFRVAnFRVDerivationLaggedVersionDerivation8}
\lambda^{\mathrm{n.n.} ( t )}_{\dot{a}} = \cos \frac{\pi \dot{a}}{t + 1} , \qquad \textrm{for} \qquad \dot{a} = 1 , \ldots , t .
\end{equation}
The formula for the Green's function (\ref{eq:GreenFunctionForTheSymmetrizedDelayMatrix}) is then immediately recovered.


\section{Examples}
\label{s:Examples}


\subsection{An Exponentially Decaying Auto--Covariance}
\label{ss:AnExponentiallyDecayingAutoCovariance}


\subsubsection{Introduction: Infinite Translationally--Invariant Matrices}
\label{sss:IntroductionInfiniteTranslationallyInvariantMatrices}

In this subsection, we will choose a particular model for the underlying auto--covariance matrix $\mathbf{A}$. It will have one generic feature, ``translational invariance,'' which means that the value of a matrix element depends only on the distance between its indices, and not on their separate values,
\begin{equation}\label{eq:TranslationalInvarianceDefinition}
A_{a b} = A ( a - b ) .
\end{equation}
Such a dependence is natural for a matrix describing temporal correlations between measurements. Moreover, we will consider $\mathbf{A}$ to be infinite, such that its indices range over both positive and negative values, $a , b \in \mathbb{Z}$.

There exists a convenient framework for dealing with infinite matrices (not necessarily fulfilling (\ref{eq:TranslationalInvarianceDefinition})): it is to perform the Fourier transformation of the matrix' indices $a , b$, replacing them in this way with continuous variables $p , q \in [ - \pi , \pi )$,
\begin{equation}\label{eq:FourierTransformationOfTheMatrixIndices}
\hat{A} ( p , q ) \equiv \sum_{a , b \in \mathbb{Z}} \ee^{\ii ( a p - b q )} A_{a b} , \qquad \textrm{or conversely,} \qquad A_{a b} = \frac{1}{4 \pi^{2}} \int_{- \pi}^{\pi} \int_{- \pi}^{\pi} \dd p \dd q \ee^{- \ii ( a p - b q )} \hat{A} ( p , q ) .
\end{equation}
For instance, the Kronecker delta \smash{$\delta_{a b}$} is mapped to the Dirac delta $2 \pi \delta ( p - q )$, and matrix multiplication is translated to the integration \smash{$\frac{1}{2 \pi} \int_{- \pi}^{\pi} \dd p ( \ldots )$}. In particular, the Fourier transform (\ref{eq:FourierTransformationOfTheMatrixIndices}) of a translationally--invariant (\ref{eq:TranslationalInvarianceDefinition}) matrix is proportional to the Dirac delta and reads
\begin{equation}\label{eq:FourierTransformationOfATranslationallyInvariant}
\hat{A} ( p , q ) = 2 \pi \delta ( p - q ) \hat{A} ( p ) , \qquad \textrm{where} \qquad \hat{A} ( p ) \equiv \sum_{d \in \mathbb{Z}} \ee^{\ii d p} A ( d ) \qquad \textrm{or conversely,} \qquad A ( d ) = \frac{1}{2 \pi} \int_{- \pi}^{\pi} \dd p \ee^{- \ii d p} \hat{A} ( p ) .
\end{equation}

Knowing the Fourier transform \smash{$\hat{A} ( p )$} allows to evaluate the moments' generating function of the matrix $\mathbf{A}$. Indeed, consider the matrix \smash{$\mathbf{G}_{\mathbf{A}} \equiv 1 / ( z \mathbf{1}_{T} - \mathbf{A} )$}. In other words, \smash{$\mathbf{G}_{\mathbf{A}} ( z \mathbf{1}_{T} - \mathbf{A} ) = \mathbf{1}_{T}$}. After the Fourier transformation, this equation can be solved as follows, \smash{$\hat{G}_{\mathbf{A}} ( p , q ) = 2 \pi \delta ( p - q ) / ( z - \hat{A} ( q ) )$}. Transforming back,
\begin{equation}\label{eq:GreenFunctionForATranslationallyInvariant1}
[ \mathbf{G}_{\mathbf{A}} ]_{a b} = \frac{1}{2 \pi} \int_{- \pi}^{\pi} \dd p \ee^{- \ii p ( a - b )} \frac{1}{z - \hat{A} ( p )} \equiv [ \mathbf{G}_{\mathbf{A}} ] ( a - b ) ,
\end{equation}
and taking trace yields the Green's function of $\mathbf{A}$,
\begin{equation}\label{eq:GreenFunctionForATranslationallyInvariant2}
G_{\mathbf{A}} ( z ) = \frac{1}{T} \Tr \mathbf{G}_{\mathbf{A}} = [ \mathbf{G}_{\mathbf{A}} ] ( 0 ) = \frac{1}{2 \pi} \int_{- \pi}^{\pi} \dd p \frac{1}{z - \hat{A} ( p )} ,
\end{equation}
where we made use of the property \smash{$( 1 / T ) \Tr \mathbf{B} = B ( 0 )$}, true for a translationally--invariant $T \times T$ ($T \to \infty$) matrix $\mathbf{B}$. Finally,
\begin{equation}\label{eq:MomentsGeneratingFunctionForATranslationallyInvariant}
M_{\mathbf{A}} ( z ) = \frac{1}{2 \pi} \int_{- \pi}^{\pi} \dd p \frac{\hat{A} ( p )}{z - \hat{A} ( p )} .
\end{equation}


\subsubsection{An Exponentially Decaying $\mathbf{A}$ and an Arbitrary $\mathbf{C}$}
\label{sss:AnExponentiallyDecayingAAndAnArbitraryC}

Let us now assume a particular translationally--invariant model of temporal covariances, namely, an exponential decay,
\begin{equation}\label{eq:AExponentiallyDecayingDefinition}
A_{a b} = A ( a - b ) \equiv \ee^{- | a - b | / \tau} ,
\end{equation}
where $\tau$ is a correlation time (in the units of the elementary time step $\delta t$), and it will be convenient to denote $\gamma \equiv \coth ( 1 / \tau )$. It is a natural model, aiming for example at sketching the temporal behavior described in par.~\ref{sss:LaggedCorrelationsBetweenTheReturns}.

We start from calculating the Fourier transform (\ref{eq:FourierTransformationOfATranslationallyInvariant}) of $\mathbf{A}$,
\begin{equation}\label{eq:AExponentiallyDecayingFourierTransform}
\hat{A} ( p ) = \frac{1 - \ee^{- 2 / \tau}}{1 - 2 \ee^{- 1 / \tau} \cos p + \ee^{- 2 / \tau}} ,
\end{equation}
which leads (\ref{eq:MomentsGeneratingFunctionForATranslationallyInvariant}) to its moments' generating function and $N$--transform,
\begin{equation}\label{eq:MomentsGeneratingFunctionAExponentiallyDecaying}
M_{\mathbf{A}} ( z ) = \frac{1}{\sqrt{1 - 2 \gamma z + z^{2}}} , \qquad \textrm{hence,} \qquad N_{\mathbf{A}} ( z ) = \gamma + \sqrt{\gamma^{2} - 1 + \frac{1}{z^{2}}} .
\end{equation}

Let the true cross--covariance matrix $\mathbf{C}$ be completely arbitrary. The pertinent equation for the moments' generating function of the estimator $\mathbf{c}$, \smash{$M \equiv M_{\mathbf{c}} ( z )$}, is (\ref{eq:DoublyCorrelatedWishartEquationForMomentsGeneratingFunction}), and for the exponentially decaying $\mathbf{A}$ (\ref{eq:AExponentiallyDecayingDefinition}) it assumes the form
\begin{equation}\label{eq:DoublyCorrelatedWishartEquationForMomentsGeneratingFunctionAExponentiallyDecayingCArbitrary}
M = M_{\mathbf{C}} \left( \frac{z}{r \gamma M + \sqrt{r^{2} \left( \gamma^{2} - 1 \right) M^{2} + 1}} \right) .
\end{equation}
For example, if \smash{$\mathbf{C} = \mathbf{1}_{N}$}, (\ref{eq:DoublyCorrelatedWishartEquationForMomentsGeneratingFunctionAExponentiallyDecayingCArbitrary}) becomes a fourth--order polynomial (Ferrari) equation,
\begin{equation}\label{eq:DoublyCorrelatedWishartEquationForMomentsGeneratingFunctionAExponentiallyDecayingCOne}
r^{2} M^{4} + 2 r ( r - \gamma z ) M^{3} + \left( z^{2} - 2 r \gamma z + r^{2} - 1 \right) M^{2} - 2 M - 1 = 0 .
\end{equation}
This result has been derived by diagrammatic methods in~\cite{BurdaJurkiewiczWaclaw2005-1}. It can be appropriated numerically to yield \smash{$M = M_{\mathbf{c}} ( z )$}, translated next to the Green's function (\ref{eq:MomentsGeneratingFunctionDefinition}), and finally to the mean spectral density (\ref{eq:SokhotskyFormula}), plotted in fig.~\ref{fig:ExponentialDecay}.

\begin{figure}
\includegraphics[width=8.5cm]{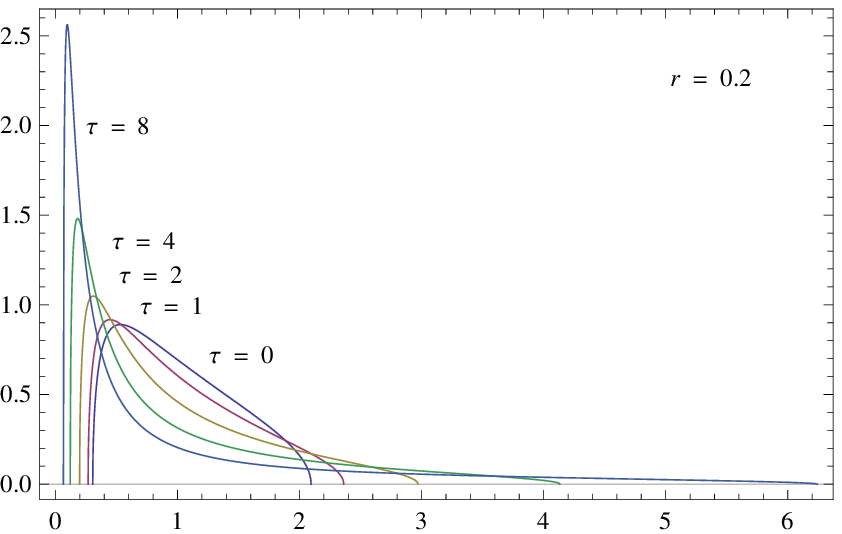}
\includegraphics[width=8.5cm]{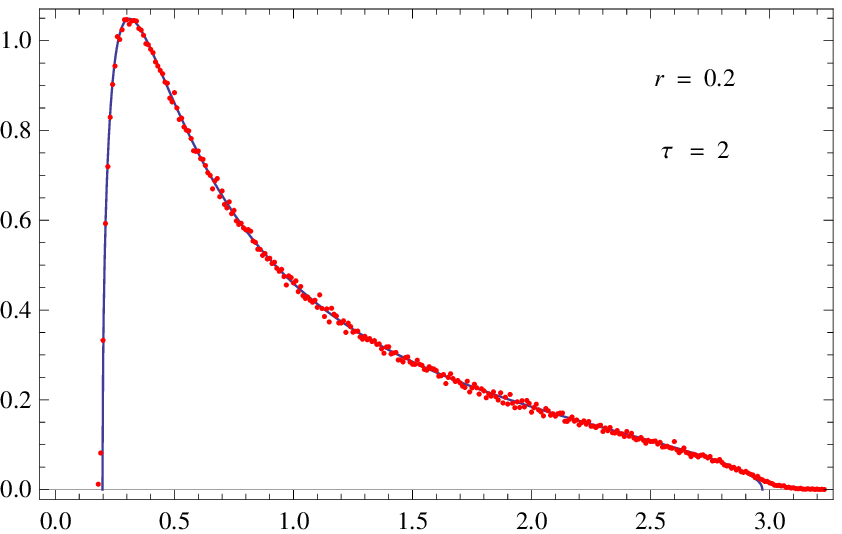}
\caption{LEFT: The theoretical eigenvalue density of the empirical cross--covariance matrix $\mathbf{c}$ for $N \to \infty$ identical normally distributed degrees of freedom, mutually uncorrelated but exponentially correlated in time (\ref{eq:AExponentiallyDecayingDefinition}), for $r = 0.2$ and $\tau = 0 , 1 , 2 , 4 , 8$.\\RIGHT: A comparison of the theoretically predicted eigenvalue density with a Monte--Carlo--generated spectrum, for $N = 100$, $r = 0.2$, $\tau = 2$, obtained by diagonalizing $4000$ matrices. Finite--size effects appear only at the edges of the spectrum.}
\label{fig:ExponentialDecay}
\end{figure}

As mentioned before, if we want to consider instead the symmetrized time--lagged estimator \smash{$\mathbf{c}^{\mathrm{sym.} ( d )}$}, in the limit (\ref{eq:ThermodynamicalLimitLaggedVersionTwo}), the resulting equation will differ from (\ref{eq:DoublyCorrelatedWishartEquationForMomentsGeneratingFunctionAExponentiallyDecayingCArbitrary}) only by formally replacing \smash{$z \to z \sqrt{1 + 2 / ( r M )}$}, so we will not print it explicitly. Then, for \smash{$\mathbf{C} = \mathbf{1}_{N}$}, an eight--order polynomial equation is found.


\subsection{The Exponentially Weighted Moving Average}
\label{ss:TheExponentiallyWeightedMovingAverage}

As extensively explained in par.~\ref{sss:Heteroscedasticity} and~\ref{sss:EstimatorsWithWeightingSchemes}, an implication of assuming that the heteroscedasticity is modeled by the I--GARCH$( 1 )$ process, is that one should use a weighted estimator for the cross--covariance matrix (\ref{eq:TransformedWeightedEstimator}) with the EWMA scheme (\ref{eq:EWMAWeights}),
\begin{equation}\label{eq:EWMADefinition}
W_{a b} \equiv T \frac{1 - \alpha}{1 - \alpha^{T}} \alpha^{a - 1} \delta_{a b} .
\end{equation}
Here $\alpha \in [ 0 , 1 ]$ is a constant, typically close to $1$ (for example, $\alpha = 0.94$ in RiskMetrics~1994~\cite{RiskMetrics1996,MinaXiao2001}), and we will actually consider the following double--scaling limit,
\begin{equation}\label{eq:EWMALimit}
N , T \to \infty , \qquad \alpha \to 1^{-} , \qquad \textrm{such that} \qquad r = \frac{N}{T} = \textrm{fixed} , \qquad \beta \equiv T ( 1 - \alpha ) = \textrm{fixed} .
\end{equation}
In this limit, the range of the EWMA suppression $\tau = - 1 / \log \alpha \sim 1 / ( 1 - \alpha ) \sim T$. (Note that the definition of $\beta$ differs from the one used in~\cite{PottersBouchaudLaloux2005}, and is more natural for relating the time cutoff $\tau$ to the length of the time series $T$.)

The moments' generating function of $\mathbf{W}$ can be explicitly calculated in the limit (\ref{eq:EWMALimit}),
$$
M_{\mathbf{W}} ( z ) = \frac{1}{T} \sum_{a = 1}^{T} \frac{1}{\frac{z}{T w_{a}} - 1} = \frac{1}{T} \sum_{a = 1}^{T} \frac{1}{\frac{z \left( 1 - \alpha^{T} \right)} {T ( 1 - \alpha ) \alpha^{a - 1}} - 1} = - \sum_{n \geq 1} \frac{z^{n} \left( 1 - \alpha^{T} \right)^{n}}{T^{n + 1} ( 1 - \alpha )^{n}} \sum_{a^{\prime} = 0}^{T - 1} \alpha^{- a^{\prime} n} - 1 =
$$
$$
= \sum_{n \geq 1} \frac{z^{n} \left( 1 - \alpha^{T} \right)^{n} \left( 1 - \alpha^{- n T} \right)}{T^{n + 1} \left( 1 - \alpha \right)^{n} \left( \alpha^{- n} - 1 \right)} - 1 = \sum_{n \geq 1} \frac{z^{n} \left( 1 - \left( 1 - \frac{r \beta}{N} \right)^{\frac{N}{r}} \right)^{n} \left( 1 - \left( 1 - \frac{r \beta}{N} \right)^{- \frac{n N}{r}} \right)}{\frac{\beta^{n}}{r} N \left( \left( 1 - \frac{\beta}{N} \right)^{- n} - 1 \right)} - 1 = \ldots ,
$$
which for $N \to \infty$ (capturing the double--scaling limit after everything has been expressed through $N$, $r$, $\beta$) simplifies to
\begin{equation}\label{eq:EWMAMomentsGeneratingFunction}
\ldots = \sum_{n \geq 1} \frac{z^{n} \left( 1 - \ee^{- \beta} \right)^{n} \left( 1 - \ee^{\beta n} \right)}{\beta^{n + 1} n} - 1 = - 1 + \frac{1}{\beta} \log \frac{1 - \frac{1}{\beta} \left( \ee^{\beta} - 1 \right) z}{1 - \frac{1}{\beta} \left( 1 - \ee^{- \beta} \right) z} .
\end{equation}
Inverting it functionally yields the $N$--transform,
\begin{equation}\label{eq:EWMANTransform}
N_{\mathbf{W}} ( z ) = \beta \frac{\ee^{\beta ( 1 + z )} - 1}{\left( \ee^{\beta} - 1 \right) \left( \ee^{\beta z} - 1 \right)} .
\end{equation}

Let us assume that the underlying cross--covariance matrix $\mathbf{C}$ is arbitrary, and there are no auto--covariances, \smash{$\mathbf{A} = \mathbf{1}_{T}$}. Then, the weighted estimator \smash{$\mathbf{c}^{\mathrm{EWMA}}$} (\ref{eq:TransformedWeightedEstimator}) is the doubly correlated Wishart random matrix with the covariance matrices $\mathbf{C}$ and $\mathbf{W}$, respectively. Having derived the $N$--transform of the latter (\ref{eq:EWMANTransform}), we may write equation (\ref{eq:DoublyCorrelatedWishartEquationForMomentsGeneratingFunction}) for the moments' generating function \smash{$M \equiv M_{\mathbf{c}^{\mathrm{EWMA}}} ( z )$},
\begin{equation}\label{eq:DoublyCorrelatedWishartEquationForMomentsGeneratingFunctionAEWMACArbitrary}
M = M_{\mathbf{C}} \left( \frac{z \left( \ee^{\beta} - 1 \right) \left( \ee^{\beta r M} - 1 \right)}{r \beta M \left( \ee^{\beta ( 1 + r M )} - 1 \right)} \right) .
\end{equation}

In particular, for \smash{$\mathbf{C} = \mathbf{1}_{N}$}, this acquires the form
\begin{equation}\label{eq:DoublyCorrelatedWishartEquationForMomentsGeneratingFunctionAEWMACOne}
z = \frac{r \beta ( 1 + M ) \left( \ee^{\beta ( 1 + r M )} - 1 \right)}{\left( \ee^{\beta} - 1 \right) \left( \ee^{\beta r M} - 1 \right)} .
\end{equation}
It is an entangled equation. In fig.~\ref{fig:EWMA1}, we present its numerical solution for $r = 0.2$ and different values of $\beta$: Since a general effect of exponential weighting is to effectively reduce the length of the sample (short memory), hence, increasing $\beta$ amounts to increasing the noise--to--signal ratio $r$, which results in broadening of the spectrum.

\begin{figure}
\centerline{\scalebox{0.7}{\rotatebox{270}{\includegraphics{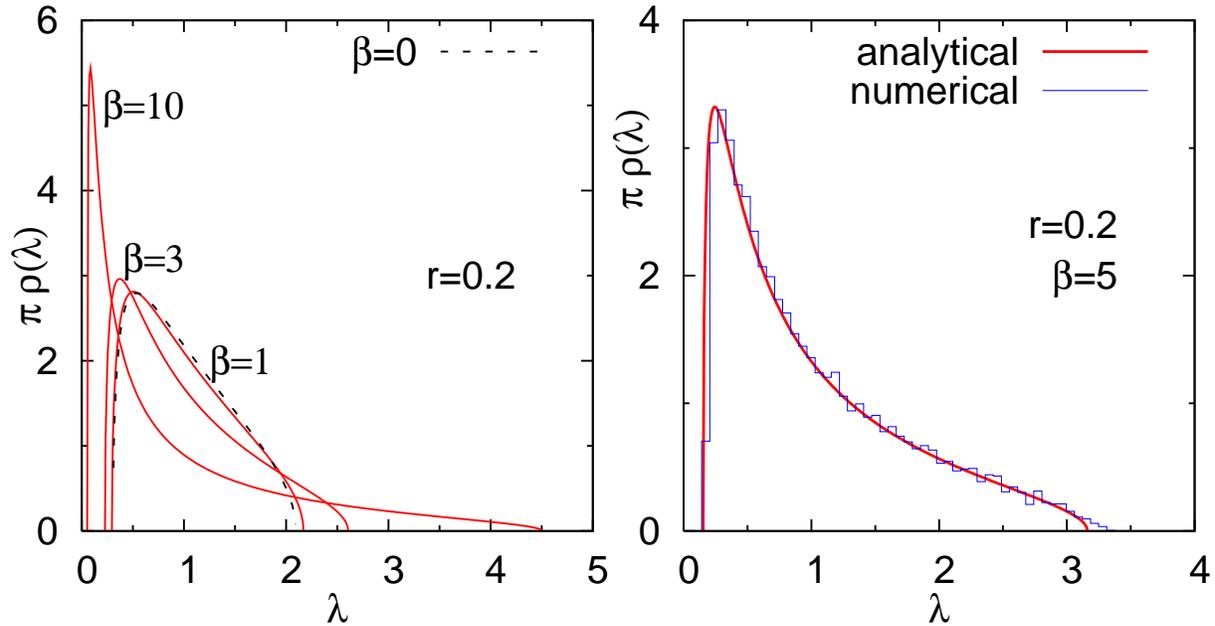}}}}
\caption{LEFT: The spectral density for $r = 0.2$ and $\beta = 0 , 1 , 3 , 10$.\\RIGHT: Comparison of the theoretical prediction for the mean spectral density from (\ref{eq:DoublyCorrelatedWishartEquationForMomentsGeneratingFunctionAEWMACOne}) with numerical calculations, for $r = 0.2$, $\beta = 5$, with $N = 100$, $T = 500$, $\alpha = 0.99$, over $1000$ samples.}
\label{fig:EWMA1}
\end{figure}

In order to compare (\ref{eq:DoublyCorrelatedWishartEquationForMomentsGeneratingFunctionAEWMACOne}) with the results known from literature, we rewrite it as an expression for the Blue's function (\ref{eq:BlueFunctionAndNTransformDefinition}) of $\mathbf{c}$,
\begin{equation}\label{eq:DoublyCorrelatedWishartBlueFunctionAEWMACOne}
B_{\mathbf{c}} ( z ) = \frac{1}{z} \left( 1 - \frac{1}{r \beta} \log \left( 1 - \frac{r \beta z}{1 - \frac{r \beta z}{\ee^{\beta} - 1}} \right) \right) ,
\end{equation}
which is slightly more general than the one recently obtained in~\cite{PafkaPottersKondor2004,PottersBouchaudLaloux2005}, where the authors first took the limit $r \to 0$, before taking the double--scaling limit (\ref{eq:EWMALimit}), and here we have an arbitrary $r$. Having a finite $r$ allows us to consistently consider the limit $\beta \to 0$, which reproduces the Mar\v{c}enko--Pastur spectrum,
\begin{equation}\label{eq:DoublyCorrelatedWishartBlueFunctionAEWMACOneBetaToZero}
B_{\mathbf{c}} ( z ) \to \frac{1}{z} + \frac{1}{1 - r z} , \qquad \textrm{as} \qquad \beta \to 0 .
\end{equation}
The limit $r \to 0$ also exists, defining the pole of the Green's function at $1$, independently of $\beta$. To recover the findings of~\cite{PafkaPottersKondor2004,PottersBouchaudLaloux2005}, we define $q \equiv r \beta$ and take $r \to 0$ with $q$ fixed,
\begin{equation}\label{eq:DoublyCorrelatedWishartBlueFunctionAEWMACOnerToZeroqFixed}
B_{\mathbf{c}} ( z ) \to \frac{1}{z} \left( 1 - \frac{1}{q} \log ( 1 - q z ) - \frac{\ee^{- q / r}}{q} ( 1 + q z ) \right) , \qquad \textrm{as} \qquad r \to 0 , \qquad q = \textrm{fixed} ,
\end{equation}
with the first two terms reproducing~\cite{PafkaPottersKondor2004,PottersBouchaudLaloux2005}, and the third term converging exponentially fast toward zero. Equation (\ref{eq:DoublyCorrelatedWishartBlueFunctionAEWMACOne}) is also useful to obtain the support of the underlying spectrum; following~\cite{Zee1996,JanikNowakPappZahed1997}, the endpoints are defined as
\begin{equation}\label{eq:EndpointsFromBlueFunction}
x_{*} = B_{\mathbf{c}} \left( z_{*} \right) , \qquad \textrm{where} \qquad B_{\mathbf{c}}^{\prime} \left( z_{*} \right) = 0 .
\end{equation}
The problem, for generic $r$ and $\beta$, may only be solved numerically; the dependence of the upper and lower endpoints of the support is shown in fig~\ref{fig:EWMA2}.

\begin{figure}
\centerline{\scalebox{0.5}{\rotatebox{270}{\includegraphics{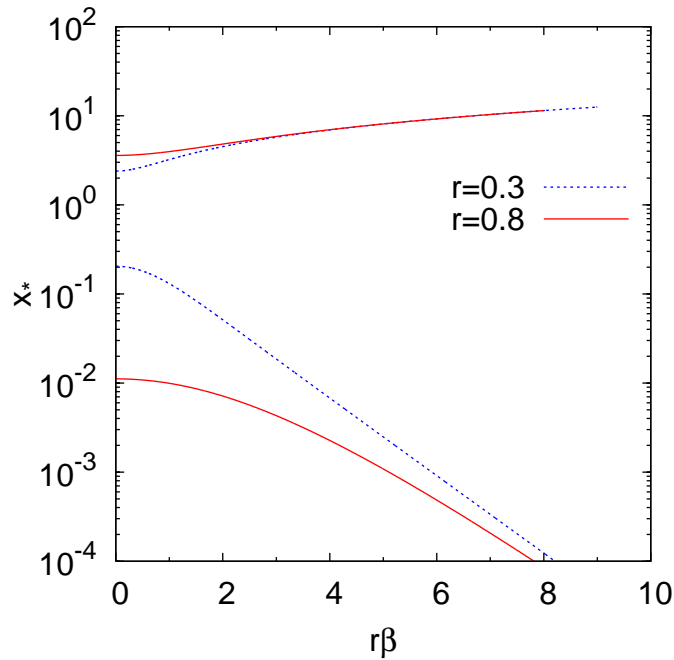}}}}
\caption{The endpoints of the support of the mean spectrum as a function of $\beta$, for $r = 0.3 , 0.8$.}
\label{fig:EWMA2}
\end{figure}


\section{Conclusions}
\label{s:Conclusions}

Equation (\ref{eq:DoublyCorrelatedWishartEquationForMomentsGeneratingFunction}) generalizes the standard result for the eigenvalue density of large--dimensional empirical covariance matrices to the case of simultaneous ``vertical'' and ``horizontal'' covariances. This set--up, however simplified, does capture many real--life problems. For example, the ``vertical'' covariances can be interpreted as between degrees of freedom present on financial markets, and the ``horizontal'' ones as temporal between the measured samples. Equation (\ref{eq:DoublyCorrelatedWishartEquationForMomentsGeneratingFunction}) provides an elegant solution to the task of estimation of the covariance matrix from historical financial time series, in a variety of ways (Pearson, time--delayed, weighted) and under diverse circumstances (with temporal correlations between the volatilities or between the residual returns, with nontrivial inter--asset correlations). Furthermore, our method has also a natural interpretation in terms of information theory, for the multiple--input--multiple--output (MIMO) systems in wireless telecommunication, where the ``vertical'' correlations are in the input and the ``horizontal'' ones in the output. Being very general and natural, this description is certainly extendable to other problems in more than few areas of science.

We attempted to introduce to the reader, in a pedagogical and practical fashion, a powerful machinery of the free random variables calculus, which is a non--commutative version of classical probability theory. It ushers in tools, based on the notion of freeness (which is non--commutative independence), which are fit to handle, in a purely algebraic and remarkably simple way, the setting of multivariate data displaying both mentioned types of covariances, without any recourse to other better known techniques of random matrix theory. Not only so, but there exist very straightforward paths to generalize FRV to situations where many of the standard RMT methods are limited, such as of heavy--tailed multivariate distributions.

We hope that this paper will have a sizable impact on the quantitative finance community, communicating the potential which lies in FRV and encouraging its applications to modeling and analyzing of financial data under involved conditions encountered in real--world problems.


\begin{acknowledgements}
We thank A. G\"{o}rlich, R. A. Janik and B. Wac\l{}aw for many interesting discussions on the subject.

This work was supported by the Marie Curie ToK project ``COCOS,'' No.~MTKD--CT--2004--517186, the EC--RTN Network ``ENRAGE,'' No.~MRTN--CT--2004--005616, and the Polish Ministry of Science Grant No.~N~N202~229137 (2009--2012). AJ acknowledges the support of Clico Ltd.
\end{acknowledgements}

\end{document}